\documentclass{aa}  

\usepackage{graphicx,epsfig,amsmath,amssymb,color,,subcaption,stmaryrd,multirow,makecell}
\usepackage{txfonts}
\usepackage{CJK}
\bibpunct{(}{)}{;}{a}{}{,}
\begin{document}

   \title{Formation of proto-cluster: \\
   a virialized structure from gravo-turbulent collapse}

   \subtitle{II. A two-dimensional analytical model for rotating and accreting system}

   \author{
          Yueh-Ning Lee \inst{1}
          \and
          Patrick Hennebelle\inst{1,2}
          }

   \institute{Laboratoire AIM, Paris-Saclay, CEA/IRFU/SAp -- CNRS -- Universit\'{e} Paris Diderot, 91191 Gif-sur-Yvette Cedex, France\\
              \email{yueh-ning.lee@cea.fr}
         \and
             LERMA (UMR CNRS 8112), Ecole Normale Sup\'{e}rieure, 75231 Paris Cedex, France\\
             \email{patrick.hennebelle@lra.ens.fr }
             }

   \date{Received 16 December 2015; accepted 25 March 2016}

 
  \abstract
   {Most stars are born in the gaseous proto-cluster environment where the gas is reprocessed after the global collapse from the diffuse molecular cloud.
   The knowledge of this intermediate step gives more accurate constraints on star formation characteristics.}
   {We demonstrate that a virialized globally supported structure, in which star formation happens, 
   is formed out of a collapsing molecular cloud,
   and derive a mapping from the parent cloud parameters to the proto-cluster to predict its properties, 
   with a view to confront analytical calculations with observations and simulations.}
   {The virial theorem is decomposed into two dimensions to account for the rotation and the flattened geometry.
   Equilibrium is found by balancing rotation, turbulence and self-gravity, while turbulence is maintained by accretion driving and dissipates in one crossing time.
   The angular momentum and the accretion rate of the proto-cluster are estimated from the parent cloud properties.}
   {The two-dimensional virial model predicts the size and velocity dispersion given the mass of the proto-cluster and that of the parent cloud.
   The gaseous proto-clusters lie on a sequence of equilibrium with the trend $R \sim M^{0.5}$,
   with limited variations depending on the evolutionary stage, the parent cloud, and the parameters not well known like turbulence driving efficiency by accretion and the turbulence anisotropy.
   The model reproduces successfully observations and simulation results.}
   {The properties of proto-clusters follow universal relations and they can be derived from that of the parent cloud.
   The gaseous proto-cluster is an important primary stage of stellar cluster formation, and should be taken into account when studying star formation.
Using simple estimates to infer the peak position of the core mass function (CMF) we find a weak dependence on the cluster mass suggesting 
that the physical conditions inside proto-clusters may contribute to set a CMF, and by extension an IMF, that looks independent of
the environment.  }

   \keywords{Turbulence
                ISM: kinematics and dynamics --
                ISM: structure --
                ISM: clouds --
                Galaxies: star clusters: general
               }

   \maketitle


\section{Introduction}
Star formation is known to be a multi-scale, multi-physics process.
As pointed out by \citet{McKee07}, 
star formation is a hierarchical process starting from giant molecular associations or HI superclouds in the diffuse ISM, 
fragmenting into giant molecular clouds and the star-forming clumps therein.
The sequential condensation from the diffuse interstellar medium to a new born stars includes density change of orders of magnitude.
Meanwhile, it is much more affordable and comprehensive if we can disentangle this complexity to some certain degree by studying this process step by step.
As stars are often born in clusters \citep{Lada03,Allen07,McKee07},
its properties are more likely determined by the cluster environment, rather than directly inherited from the parent molecular cloud.
Understanding how the cluster forms out of a molecular cloud would thus provide us with useful information on the initial conditions of the star-forming environment,
and would give important clues to the origin of the initial mass function (IMF). 
\citet{Pfalzner16} recently pointed out that there is a tight correlation between the mass-size relations of star-forming clumps and stellar clusters, 
and suggested that cluster studies should be performed in conditions corresponding to the clumps. 

Star-forming clumps that have been observed span a broad range of mass and radius.
A cluster mass-size relation $R \propto M^{0.38}$ has been inferred from observations by \citet{Fall10} for clumps having mass from $10^2$ to $10^4$ solar masses. 
A more complete dataset from the ATLASGAL survey \citep{Urquhart14} shows the dependence $R \propto M^{0.50}$.  
The dispersions in those data are compatible with a power-law exponent between $0.4$ and $0.6$, 
leaving some uncertainties. 
Theoretical works have been done to better understand the gaseous proto-clusters, 
the gas-dominated primary phase of cluster formation. 
\citet{Hennebelle12} previously derived the mass-size relation 
by balancing the gravitational energy, the turbulent energy, and the ram pressure of an accreting system.
The relation $R \propto M^{1/2}$  or $R \propto M^{2/3}$ was found with different accretion schemes and reproduced successfully the observational results.
He then applied the Hennebelle-Chabrier theory of star formation \citep{HC08,HC09,HC13} to the gas inside the gaseous proto-cluster to obtain the cluster IMF. 
\citet{Pfalzner11} regarded this mass-size relation as a growth sequence, 
and \citet{Parmentier13} applied a star formation model which depends on local density to relate star-forming clumps to stellar clusters.

So far these  models have ignored  rotation. However rotation has been observed in several stellar clusters \citep{HB12,Davies11,Mackey13},
moreover it is not known to be a common feature, that has been found in most, if not all, contracting systems \citep{Longmore14}.
Indeed structures forming from collapse usually exhibit important rotational motions due to angular momentum conservation (see companion paper I). 
We therefore follow a similar idea as \citet{Hennebelle12}, while taking into account the rotation as well as the turbulence, to develop an analytical model to derive gaseous proto-cluster properties from those of the parent cloud out of which they form.

In the companion simulation paper (paper I), we see a virialized structure forming at the center of a collapsing molecular cloud simulation. 
The goal of the present paper is to provide an analytical model which could be confronted with the numerical results. 
This gas-dominated  structure, the proto-cluster, 
is the primary site of star formation and will evolve into a stellar cluster subsequently.
It is bound by its own gravitational potential and the ram pressure of the infalling flow,
and is supported by supersonic turbulence which is nourished by the accretion.
Observational analyses of the star-forming clump G28.34+0.06 P1 performed by \citet{Zhang15} demonstrate that the $10^3$ M$_\odot$, 0.6 pc object is close to virial equilibrium,
and that cores are forming from fragmentation while mass is being accreted from the filament in which it is embedded.
In this paper, we introduce a simple analytical model very similar to this scenario to account for the formation of the gaseous proto-cluster,
and to predict its mass-size relation. 
Observationally, the infall motion is detected with a double-peaked blue-skewed line profile, 
while complicated by density, molecular abundance, and excitation temperature. 
Besides, clear global infall at clump scale is difficult to detect \citep{Lopez10, Reiter11} since local star formation also create infall and outflow signatures.
Infall rate of star-forming clumps are observed to be small, about 10\% of the free-fall velocity \citep{Rolffs11, Tan14, Wyrowski16}, 
supporting a quasi-static picture. 
\citet{Rygl13} suggested that the infall and outflow only become evident when the clump-cloud column density contrast exceeds 2, 
and at more evolved stage the infall is halted. 

Given the small infall and the observed sequence \citep{Fall10, Urquhart14}, 
an equilibrium, at least not fastly and globally collapsing, 
assumption is highly plausible. 
Our model applies to low mass clusters, 
those with mass above $10^4$ are beyond the scope of our discussion since more massive clumps tend to evolve more quickly and form massive stars which in turn disrupt the cloud via feedback. 
In consequence, massive star-forming clumps live shorter and possibly follow a different evolutionary track. 
One motivation to propose this quasi-static equilibrium model comes from the fact that we do see them forming in collapsing molecular cloud simulations (paper I). 
On the other hand, 
the gaseous proto-cluster must live for long enough to be observed, 
and the tight correlation between their mass and size should be a consequence of certain self-regulation during the evolution.  
We should caution that although we refer to this gaseous proto-cluster as the first phase of stellar cluster formation, 
this is actually a continuous process: 
the stars start to form as the gas proto-cluster is still accreting, 
and finally take over after the gas expulsion. 
In this picture, 
the gaseous proto-cluster is globally in equilibrium while density fluctuations therein cause local infall to form stars. 
In this study, we restrict ourselves to the earliest stage where the effects of the stars are less important. 
In \S 2, we first describe how the virial theorem should be decomposed into two dimensions to account for a flattened rotating system.
The influences of model parameters are discussed in \S 3.
As sink particle are allowed to form in the simulation, 
we also adapt our analytical model to yield a comparison with the simulation results. 
These are followed by the conclusions in \S 4.


\section{Two-dimensional model: modified virial theorem}
In this section we present a simple analytical model to describe the gaseous proto-cluster formation from collapse of a molecular cloud:
a virial model, adapted to the turbulent and rotational kinematics of the cluster under mass accretion.
We propose that a gaseous proto-cluster is a structure in global virial equilibrium where the rotational and turbulent kinetic energy support against its self-gravity and ram pressure confinement. 
We stress that this global equilibrium does not preclude local infall and star formation. 
Indeed this is what we observe in the simulations. 
The important rotation results from the amplification due to angular momentum conservation under collapse,
while the turbulence is sustained by the accreting gas, and decays in one crossing time.

We start by deducing the two-dimensional virial theorem for an ellipsoidal structure with rotation and turbulence,
and then discuss the energy balance from accretion and turbulent dissipation.
The mass accretion rate and the specific angular momentum are evaluated as properties of the parent cloud.
Finally, we present the mass-size relation predicted by the model.

\subsection{Two-dimensional virial theorem of the gaseous proto-cluster}

Since a cluster with rotational motion is indeed anisotropic, 
a spherically symmetric model could be too simplistic and may fail to capture all essential features.
In this section a two-dimensional model of an oblate ellipsoid is considered,
with its minor axis coinciding with the rotational axis.
The three semi-axes are $ a = b = R > c = H$. 
Virial theorem has been discussed in tensor form by \citet{Parker54}, 
showing how forces in different directions are balanced. 
We demonstrate how the virial integral of the system is calculated in  two dimensions.

The thermal terms are neglected for simplicity because turbulence is generally supersonic, 
so is the magnetic pressure since it is not very important at the cluster scale, 
as seen in the simulations (Fig. 11 of paper I).
The model is axisymmetric and we take the inner product of the momentum equation 
$\rho d_t \vec{v}  = - \rho \nabla \phi$ 
respectively with the $\vec{r}$ and $\vec{z}$ vectors in the cylindrical coordinate before integrating over the volume of the ellipsoid.
Let us first consider the inner product with $\vec{r}$:
\begin{align} 
\int \limits_V \rho d_t \vec{v}  \cdot \vec{r} dV=& \int \limits_V - \rho \nabla \phi \cdot \vec{r}dV. 
\end{align}
The integration gives (see appendix \ref{appen_vir} for detailed derivations) 
\begin{align} 
&{1\over 2} \partial_t^2 \int \limits_V \rho r^2 dV + {1\over 2} \partial_t \int \limits_S \rho \ r^2 \vec{v} \cdot \vec{dS} +  \int \limits_S  v_r r \rho \vec{v}\cdot \vec{dS}  - \int \limits_V  \rho v_\mathrm{2d}^2 dV \nonumber\\
&= -{3\over 5} {GM^2\over R}   \left[{\cos^{-1}{(\eta)}\over (1-\eta^2)^{3\over 2}} - {\eta\over {1-\eta^2}}\right]
=-{GM^2\over R} u_r(\eta),
\label{virial_r}
\end{align}
where $v_\mathrm{2d}$ is the velocity in the $x-y$ plane, and $v_r$ is the velocity in the $\vec{r}$ direction;
$\eta = {H\over R}$ represents the aspect ratio of the ellipsoid.
In the $\vec{z}$ direction we obtain:
\begin{align} 
\int \limits_V \rho d_t \vec{v}  \cdot \vec{z} dV=& \int \limits_V - \rho \nabla \phi \cdot \vec{z}dV 
\end{align}
\begin{align} 
&{1\over 2} \partial_t^2 \int \limits_V \rho z^2 dV + {1\over 2} \partial_t \int \limits_S \rho \ z^2 \vec{v} \cdot \vec{dS} +  \int \limits_S  v_z z \rho \vec{v} \cdot \vec{dS}  - \int \limits_V  \rho v_\mathrm{1d}^2 dV \nonumber\\
&=-{3\over 5} {GM^2\over R}   \left[ {\eta \over 1-\eta^2 } -{\eta^2 \cos^{-1}{(\eta)} \over (1-\eta^2)^{3\over 2}} \right]
=-{GM^2\over R} u_z(\eta), 
\label{virial_z}
\end{align}
where $v_{1d} = v_z$ is the velocity in the $\vec{z}$ direction. 

Virial equilibrium is reached when the time derivative terms are zero.
The first term on the left hand side of Eqs. (\ref{virial_r}) and (\ref{virial_z}) is analogous to the acceleration of inertia changing rate,
and second term corresponds to the change in mass accretion rate multiplied by the surface.
As we will describe later, the mass accretion rate can be roughly regarded as a constant of time,
and therefore these terms become less important and can be treated as zero in equilibrium at later phase when the gaseous proto-cluster has gained sufficiently high mass.
The third term on the left hand side of both equations is the counterpart of the ram pressure term in the spherical model \citep{Hennebelle12},
while the geometry renders its interpretation less obvious since we are ignorant of the mass infall pattern.
There exist solutions in two regimes: gravitation dominated and ram pressure dominated.
While the ram pressure dominated solution has too large radius and too low density, 
not corresponding to the conditions discussed here and the supersonic approximation probably not valid,
it is neglected for simplicity of the discussion.
Readers are invited to refer to appendix \ref{appen_pram}  for further discussions on ram pressure
and see that the effect on cluster size is indeed a small correction.
Therefore we have the equations for virial equilibrium in two dimensions by simplifying Eqs. (\ref{virial_r}) and  (\ref{virial_z}):
\begin{align} 
Mv_\mathrm{2d}^2 &=  {GM^2\over R} u_r(\eta) \label{eqn_v1}\\
Mv_\mathrm{1d}^2 &=  {GM^2\over R} u_z(\eta) \label{eqn_v2} 
\end{align}

\subsection{The energy equilibrium of an ellipsoidal cluster}

In an accreting system, 
the turbulence is driven by the gravitational energy released from the accreted material \citep{Klessen10, Goldbaum11}, 
and it dissipates through the turbulent cascade on the time scale of the crossing time of the system.
The turbulent energy is found by balancing the energy released from accretion and its dissipation.
Studies such as that of Newton, Maclaurin, and Jacobi have been done to understand the force balancing in a homogeneous uniformly rotating body, 
and they calculated the gravitational force for oblate ellipsoids \citep{Chandrasekhar67, Binney08}. 
\citet{Neutsch79} explored related functions for ellipsoidal bodies. 
He derived the gravitational potential energies for ellipsoids with uniform, gaussian, and exponential density profiles, and found them to be different only by a factor of order unity. 
We consider an uniform ellipsoid with semi-axes $a \geq b \geq c$,
and we therefore have the gravitational potential energy given by the Legendre elliptical integral of the first kind $F(\varphi | k) = \int_0^\varphi {d\theta \over \sqrt{1-k^2 \sin{\theta}^2}}$: 
\begin{align} 
E_\mathrm{grav} &= {3 \over 10} G M^2 {2 \over \sqrt{a^2 -c^2}} F\left(\cos^{-1}{\left({c \over a}\right)} \middle| {a^2 -b^2 \over a^2 -c^2}\right) \\
&=  {3 \over 10} G M^2 {2 \over \sqrt{R^2 -H^2}} \cos^{-1}{\left({H \over R}\right)}  \nonumber\\
&= {3 \over 5}  {G M^2 \over R}{\cos^{-1}{(\eta)} \over \sqrt{1 -\eta^2}}  
=  {GM^2\over R} u_g(\eta). \nonumber
\end{align}
If we assume that the density stays uniform, 
the released gravitational energy available to aliment the turbulence is:
\begin{align} 
\dot{E}_\text{grav} =  \epsilon {GM^2\over R} u_g(\eta) \left({2\dot{M} \over M} -{\dot{R}\over R} + {u_g'(\eta)\dot{\eta}\over u_g(\eta)} \right),
\end{align}
where the unknown factor $\epsilon \leq 1$ stands for the transformation efficiency of gravitational energy into kinetic energy,
while part of the energy is dissipated at the accretion shock.
The change in $\eta$ should be small compared to that in $M$.
The change of radius is also small at the beginning of gaseous proto-cluster formation and thus is negligible. 
At later time once the stationary regime is reached, 
it should follow a power-law dependance on the mass and therefore the second term is proportional to the first term. 
By absorbing the second and third terms in the parenthesis into the uncertainties of accretion driving efficiency,
we obtain  
\begin{align} 
\dot{E}_\text{grav} = \epsilon_\mathrm{acc} {2GM\dot{M}\over R} u_g(\eta).
\label{eqn_grav} 
\end{align}

The turbulence dissipates via turbulent cascade on the crossing time of the system $\tau_\mathrm{diss}$,
while the directional energy distribution and the relevant scale is less well understood in ellipsoidal geometry.
We discuss two sets of assumptions.
Firstly, we assume that the turbulence is anisotropic and that the turbulent energy follows the inertial regime of the Kolmogorov spectrum despite that the structure is not spherically symmetric,
that is, the energy cascades down length scales at the same rate and is dissipated eventually.
This implies $\sigma_R^3/R =\sigma_H^3/H$, 
where $\sigma_R$ and $\sigma_H$ are the rms velocity of vortices of size $R$ and $H$ respectively.
The velocity dispersion $\sigma_x$ in the $x$ direction has contributions from motions parallel to the $x-y$ plane where the limiting scale is $R$ and those parallel to the $x-z$ plane with scale $H$,
and likewise for $\sigma_y$.
In the direction parallel to the short axis, $\sigma_z$ is limited both in $x-z$ and $y-z$ planes by the semi-minor axis $H$.
We thus have
\begin{subequations}
\begin{align} 
\sigma_x^2 &= \sigma_y^2 = (1+\eta^{2\over 3}) \sigma_R^2 /2 \\
\sigma_z^2 &= \eta^{2\over 3} \sigma_R^2 \\
\dot{E}_\mathrm{diss} &= {3\over 2} M \sigma_H^2 / \tau_\mathrm{diss}= {3\over 2} M \sigma_H^2 / {2H \over \sigma_H} =  {3\over 4} M \sigma_R^3 / R. 
\end{align}
\label{eqn_kol}
\end{subequations}
On the other hand, 
if the inertial range at which the energy is dominating is reached at scales smaller than the cluster size,
the turbulence should be isotropic.
By assuming the dominating scale $\epsilon_\mathrm{diss} H < H$,
we obtain
\begin{subequations}
\begin{align} 
\sigma_x^2 &= \sigma_y^2 = \sigma_z^2 = \sigma_R^2 \\
\dot{E}_\mathrm{diss} &= {3\over 2} M \sigma_R^2 / \tau_\mathrm{diss} = {3\over 2} M \sigma_R^2 / {2\epsilon_\mathrm{diss} H\over \sigma_R} = {3\over 4} M \sigma_R^3 /  \epsilon_\mathrm{diss} \eta R.
\end{align}
 \label{eqn_iso}
\end{subequations}
This introduces two sets of equations with similar form but different coefficients when we proceed to solve for equilibrium solutions.

\begin{figure}[]
\centering
\includegraphics[width=0.5\textwidth]{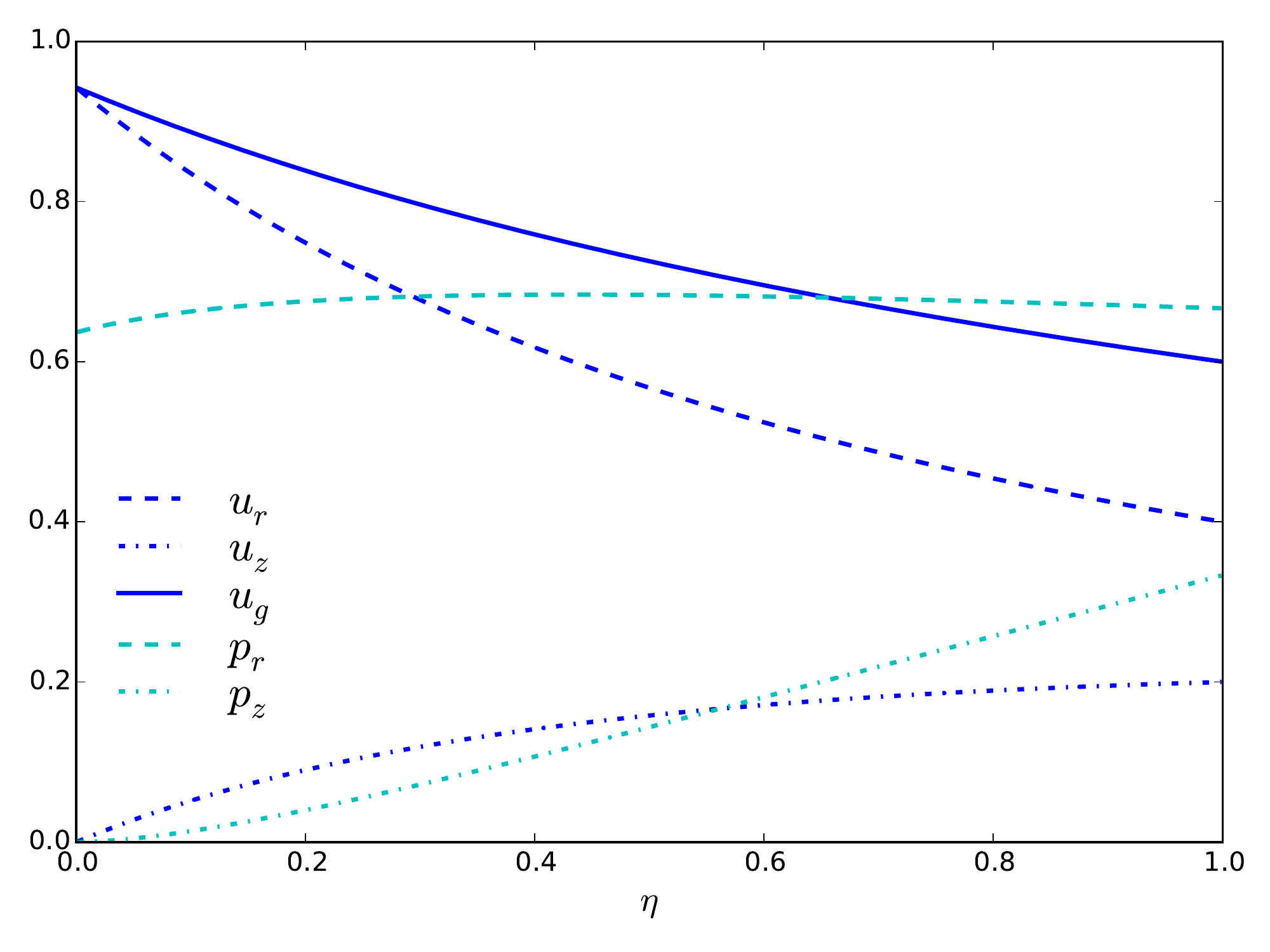}
\caption{The gravitational potential in blue curves (solid: complete potential $u_g$, dashed: $u_r$, dot-dashed: $u_z$) and the ram pressure factors in cyan (dashed: $p_r$, dot-dashed: $p_z$, see appendix \ref{appen_pram}) as functions of the ellipsoid aspect ratio $\eta$.
}
\label{uruzug}
\end{figure}

The rotation provides support only in the directions perpendicular to the rotational axis.
By using the averaged specific angular momentum $j$,
the rotational energy of the cluster is $2E_\mathrm{rot} = {5\over 2}M\left({j \over R}\right)^2$.
The factor ${5\over 2}$ comes from uniform density and rigid body assumptions,
which suffers from some uncertainty since we do not know the actual distribution of mass and angular momentum.
A numerical evaluation from simulation results is done by taking the ratio between 
$JI^{-1}J$, 
where $J$ and $I$ are the angular momentum and the rotational inertia matrix of the cluster, 
and $M\left({j \over R}\right)^2$ since the former is a good estimation of the rotational energy.
This gives values ranging between 2 and 4, confirming that we are not too far from reality.
Finally, by using Eqs. (\ref{eqn_v1}, \ref{eqn_v2}, \ref{eqn_grav}, \ref{eqn_kol}), and (\ref{eqn_iso}) while splitting the two dimensional motion perpendicular to the short axis into rotation and turbulence,
we obtain the equation set to be solved:
\begin{subequations}
\label{eqnset}
\begin{align} 
{5\over 2}\left({j \over R}\right)^2 + s_r(\eta) \sigma^2 &= {GM \over R} u_r(\eta) \\
s_z(\eta) \sigma^2 &= {GM \over R} u_z(\eta)  \\
\dot{E}_\mathrm{diss} / M= d(\eta) { \sigma^3 \over  4 R  } &= \dot{E}_\mathrm{grav} / M= \epsilon_\mathrm{acc} {2G\dot{M} \over R} u_g(\eta), 
\end{align}
\end{subequations}
where the geometrical factors $s_r$, $s_z$, and $d$ are described in Eqs. (\ref{eqn_kol}) and (\ref{eqn_iso}) for two cases.
The subscript of $\sigma_R$ is omitted for simplicity.
The factors $u_r$, $u_z$, and $u_g$ are functions of the aspect ratio $\eta = {H\over R}$ of the ellipsoid, 
as shown in Fig. \ref{uruzug}. 
The essential idea is to decompose the gravitational potential resulting from force in different directions.  
It could be readily verified that $u_r(\eta)+u_z(\eta)=u_g(\eta)$,
and that $u_g(1)={3\over5}$ corresponds to the spherical case.

\subsection{Accretion rate}
The solutions of Eqs. (\ref{eqnset}) 
are set by the mass, accretion rate, and the specific angular momentum,
The latter two are estimated as functions of the cluster mass, or more precisely, the cloud mass.

The accretion rate is estimated with the free-fall collapse of the parent cloud,
while assuming that the accretion rate of a cluster inside a cloud is close to that of the cloud if the masses are comparable. 
We start by calculating the time of free-fall into a finite volume, 
and show that the mass accretion rate reaches almost a constant when the central mass exceeds about a tenth of the cloud mass. 
The cloud density profile used in our simulations (see Paper I) and its mean density inside radius $r$ are
\begin{align}
\rho(r) &= {\rho_0 \over 1+ \xi^2}, ~\xi={r\over r_0}\in [0,\xi_\mathrm{ext}] \label{den_pro}\\
\overline{\rho}(r) &= { \int_{V(r)} \rho(r^\prime) dV(r^\prime) \over  V(r) } = 3\rho_0 \left(\xi-\tan^{-1}{\xi}\right)\xi^{-3},
\end{align}
where $\xi_\mathrm{ext} = r_\mathrm{ext} / r_0 = 3$ in our case,
and we consider the free-fall time for a shell mass at $r$ to arrive at the cluster radius $r_f$ (see appendix \ref{appen_ff} for detailed derivation):
\begin{align}
t_\mathrm{ff}(r) &= \sqrt{{3\over 8\pi G \overline{\rho}(r)}} \left[\cos^{-1}{\left(\sqrt{{r_f\over r}}\right)}+\sqrt{{r_f\over r}\left(1-{r_f\over r}\right)} \right].
\label{tffr}
\end{align}
It could be alternatively written with the normalized parameters $\xi$ and $\xi_f = r_f/r_0$:
\begin{align}
t_\mathrm{ff}(\xi) &= \sqrt{{1\over 8\pi G \rho_0}} \left(\xi-\tan^{-1}{\xi}\right)^{-1\over 2}\xi^{3\over 2} \times \nonumber \\
& \left[\cos^{-1}{\left(\sqrt{{\xi_f\over \xi}}\right)}+\sqrt{{\xi_f\over \xi}\left(1-{\xi_f\over \xi}\right)} \right].
\end{align}
Consider a cloud which starts as static and collapses in free-fall,
the collapse proceeds as an outward expansion wave from the center.
This equation signifies that at $t_\mathrm{ff}(r)$ all the mass originally inside radius $r$ is accreted onto the cluster.
We therefore have the mass inside the cluster radius $r_f$ at time $t$:
\begin{align}
M(t) &= M(t_\mathrm{ff}^{-1}(t)) = 4 \pi \rho_0 r_0^3 \left(\xi-\tan^{-1}\left({\xi}\right)\right), ~ t_\mathrm{ff}(\xi) = t 
\end{align}
The mass accretion is represented as the rate each new shell mass is included:
\begin{align}
\label{Mdot0t}
\dot{M}_0(t) &=  \dot{M}(t_\mathrm{ff}^{-1}(t)) = 4\pi r(t)^2 \rho(r(t)) {dr(t) \over dt} \\
&=  4\pi r(t)^2 \rho(r(t)) ~/~ {dt_\mathrm{ff} \over dr} \nonumber \\
&=  4\pi r_0^2 \xi^2  {\rho_0 \over 1+ \xi^2} r_0 ~/ ~{dt_\mathrm{ff} \over d\xi}\nonumber \\
&=  {M_\mathrm{c} \sqrt{8\pi G \rho_0} \over \left(\xi_\mathrm{ext}-\tan^{-1}{\left(\xi_\mathrm{ext}\right)}\right)  }
\times  \xi^{3\over 2}\left(\xi-\tan^{-1}{\left(\xi\right)}\right)^{1\over 2}  \times \nonumber\\ 
&\left\{ \left[{-\xi^3/2 \over {\xi-\tan^{-1}{\xi}}}+{3\over 2}(1+\xi^2)\right]\left[\cos^{-1}{\left(\sqrt{{\xi_f\over \xi}}\right)}+\sqrt{{\xi_f\over \xi}\left(1-{\xi_f\over \xi}\right)} \right]\right. \nonumber \\
& \left. + (1+\xi^2){\xi_f\over \xi} \sqrt{{\xi_f\over \xi} \over 1-{\xi_f\over \xi}} \right\}^{-1} \nonumber 
\end{align}
\begin{figure}[]
\centering
\includegraphics[width=0.5\textwidth]{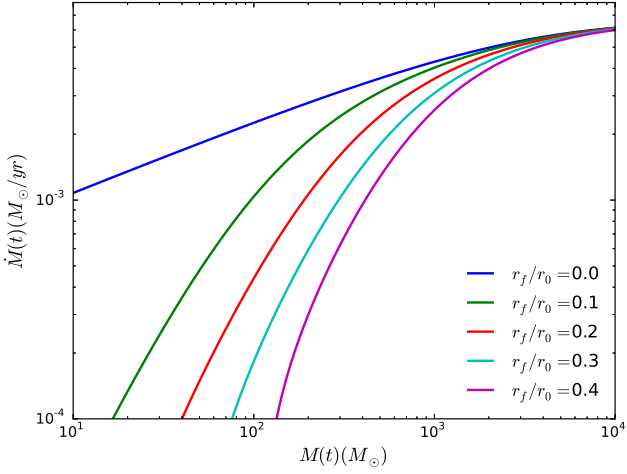}
\caption{The mass accretion rate plotted against mass for several final radius values $r_f/r_0$ inside a $10^4$ M$_\odot$ cloud,
assuming the interior of the cloud collapses in free-fall.
The radius at which $\dot{M}$ is evaluated increases from the top to the bottom curve.}
\label{MdotMplot}
\end{figure}

\begin{figure}[]
\centering
\includegraphics[width=0.5\textwidth]{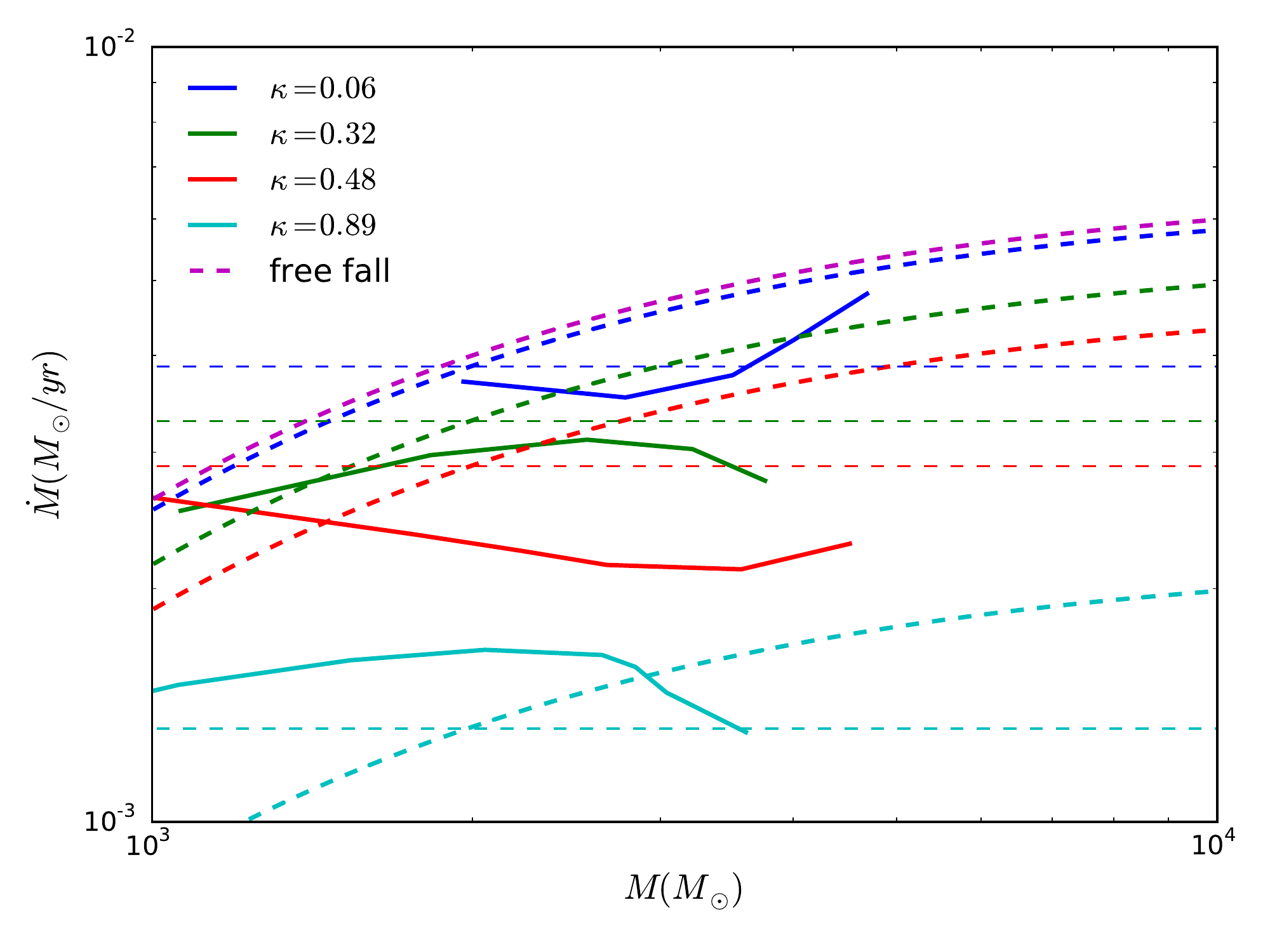}
\caption{The mass accretion rate plotted against mass at $r=1$ pc. 
The solid curves represent values evaluated in simulations with different levels of turbulent support,
where $\kappa$ is the ratio of turbulent over gravitational energy.
The dashed curve in magenta is the analytical solution for a $10^4$ M$_\odot$ cloud in free fall.
The accretion rate with the empirical correction $\sqrt{1-\kappa}$ is plotted in dashed curves with corresponding colors.
The accretion rate is approximated to be constant at its value at $M_\ast = 2 \times 10^{3}$ M$_\odot$.}
\label{Mdot_simu}
\end{figure}

For a given $\xi_f$, 
the mass $M(r_f)$ and the accretion rate $\dot{M}(r_f)$ of the gaseous proto-cluster could be calculated at the same time as functions of $\xi$.
Figure \ref{MdotMplot} displays the gaseous proto-cluster mass accretion rate plotted against mass in a molecular cloud of $10^4$ solar mass at several final radii $r_f$, 
which correspond to the range of the proto-cluster radius in our simulations. 
After the gaseous proto-cluster mass reaches over $10^3$ M$_\odot$, 
the accretion rate becomes weakly dependent of the cluster mass for any cluster radius and approaches a value characteristic of the cloud mass.
It is therefore reasonable to estimate the cluster accretion rate with the cloud mass and from Eq. (\ref{Mdot0t}) we obtain
\begin{align} 
\dot{M}_0 &\propto  M_\mathrm{c} \sqrt{\rho_\mathrm{c}},
\end{align}
where $M_\mathrm{c}$ and $\rho_\mathrm{c}$ are the cloud mass and density.
Consider a molecular cloud following the Larson relations \citep{Larson81,Falgarone04, Falgarone09,HF12, Lombardi10}:
\begin{align}
\rho_\mathrm{c} \propto R_\mathrm{c}^{-\gamma}  ~~~ \mathrm{and}~~~~
\sigma_\mathrm{c} \propto \sqrt{M\over R} \propto R_\mathrm{c}^{1-{\gamma \over 2}},
\label{Larson}
\end{align}
where different values of $\gamma$ (typically around 0.7 or 1) are being quoted in the literature. 
We obtain
\begin{align} 
\label{mdot_m}
\dot{M}_0 &\propto  M_\mathrm{c} \sqrt{\rho_\mathrm{c}}  \propto R_\mathrm{c}^3  \;\rho_\mathrm{c}^{1.5}
\sim R_\mathrm{c}^{3-1.5\gamma} \propto M_c^{(6-3\gamma)/(6-2\gamma)}.
\end{align}
With $\gamma = 1$ or $0.7$, we get $\dot{M}_0 \propto M_c^{\gamma_{\dot{M}}}$ where $\gamma_{\dot{M}} = 0.75$ or $0.85$ respectively.
While the accretion rate $\dot{M}_0$ is estimated for a cloud without turbulent support,
an empirical correction is made by multiplying the self-gravitating force by a dilution factor $(1-\kappa)$,
where $\kappa = E_\mathrm{turb} / E_\mathrm{grav}$ is the ratio between turbulent kinetic energy and gravitational energy of the cloud.
This intervenes in the accretion rate through Eq. (\ref{tffr}) and lengthens the free-fall time by a factor $1/\sqrt{1-\kappa}$.
A numerical evaluation is made for the cloud of $10^4$ solar mass as that in our simulations, obtaining
\begin{align} 
\label{eq_mdot}
\dot{M} &= \dot{M}_0 \sqrt{1-\kappa} \\
&= 4.0 \times 10^{-3} ~\mathrm{M}_\odot ~ \mathrm{yr}^{-1} \left({\alpha_{\ast,c}M_\ast \over 10^4 \mathrm{ M}_\odot}\right)^{\gamma_{\dot{M}}} 
\sqrt{1-0.35\left({\sigma_\mathrm{rms} \over \sigma_\mathrm{vir}}\right)^2}, \nonumber
\end{align}
where $\alpha_{\ast,\mathrm{c}} = M_\mathrm{c} / M_\ast$ is the cloud-cluster mass ratio.
The $\kappa$ factor scales to the square of the ratio between the actual velocity dispersion and that of a virialized cloud.
A geometrical factor $0.35$ accounts for the centrally concentrated mass distribution and accords the estimations to the values in simulations.
Simplifying by using the same accretion rate for proto-cluster of any mass inside the cloud of same mass, 
the canonical value $4.0 \times 10^{-3}$ M$_\odot ~ \mathrm{yr}^{-1}$ which corresponds to the accretion rate at $M_\ast = 2 \times 10^{3}$ M$_\odot$ is used 
as normalization reference. 
We stress that the parameter $\alpha_{\ast,\mathrm{c}}$ is setting the strength of the accretion rate. 
Lower values of $\alpha_{\ast,\mathrm{c}}$ simply corresponds to lower accretion.  
An example of mass accretion rate evaluated at radius 1 pc,
corresponding to $r_f/r_0 = 0.13$, is shown in Fig. \ref{Mdot_simu} for four simulations with different levels of turbulent support.
The mass accretion rate in the simulations is evaluated as follows:
We first calculate the mass contained inside an ellipsoidal region (defined in paper I) which has the same volume as that of a 1 pc radius sphere at several time steps.
The mass is then fitted as a function of time and in turn gives its time derivative.
The accretion rates evaluated from simulations are plotted with solid curves, 
and the modeled values multiplied by the correction factors with dashed lines. 
The model value is taken at $M_\ast = 2 \times 10^{3}$ M$_\odot$ and the accretion rate is assumed to be constant for all cluster masses (only function of cloud mass). 
This empirical correction for turbulent support is very simplistic. 
However, the error is within a factor 2, 
and the four turbulence levels cover a large range of accretion rate which all give reasonable mass-size relations (see \S \ref{sec_m_s}). 
Thus we conclude that this accretion rate approximation does not affect too much the results.
Observations show that star-forming clumps accrete at about 10$^{-3}$ M$_\odot$ yr$^{-1}$ with an infall velocity around 1 km s$^{-1}$ \citep{Fuller05, Peretto06, Lopez10, Rygl13}, 
which is coherent with our model estimation.

\subsection{Angular momentum}
\begin{figure}[]
\centering
\includegraphics[width=0.5\textwidth]{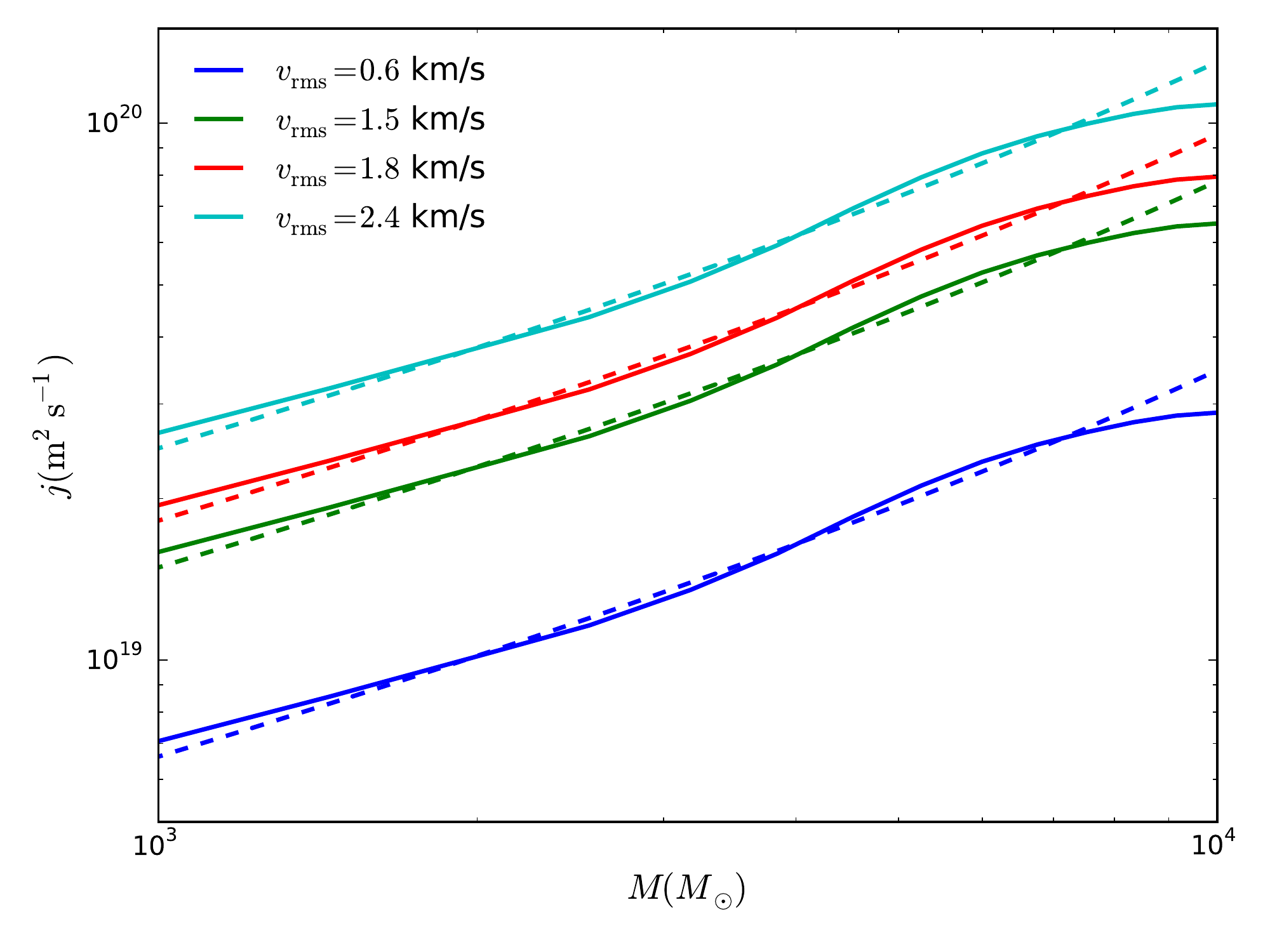}
\caption{The specific angular momentum plotted against mass contained inside varying radius.
Solid curves represent the values calculated in ellipsoidal regions (described in paper I) inside the initial cloud.
Dashed curves show the values estimated analytically. 
Values from four simulation with various initial turbulent supports are shown, 
and the specific angular momentum scales with the velocity dispersion.}
\label{j_simu}
\end{figure}

The estimation of angular momentum is motivated by the simulation.
In a large molecular cloud, there exists very often a residual rotation after cancelation of turbulent vortices,
and this rotation becomes important as the cloud collapses due to angular momentum conservation, 
as demonstrated by the remarkable rotation of the gaseous proto-cluster (paper I).
For a simple analysis, we take a characteristic rotational velocity proportional to the turbulent velocity dispersion
and following the scaling law $v_\mathrm{rot} \propto v_\mathrm{rms} (r/r_0)^{0.5}$ \citep{Dib10,Burkert00}.
By assuming random rotational axis in spherical shells, 
we have the averaged specific angular momentum inside a sphere of radius $r$ with the density profile described in Eq. (\ref{den_pro}):
\begin{align} 
\overline{j}(r) &= \left[  \frac{\int \limits_{M(r)}(v_{rot} r)^2 dm}{\int  \limits_{M(r)} dm} \right]^{1\over 2} \\
&=0.22~ v_\mathrm{rms} ~r_0 ~{1 \over 2} \left[  \frac{\xi^4-2\xi^2 + 2\log{(1+\xi^2)} }{\xi-\arctan{\left(\xi\right)}} \right]^{1\over 2},\nonumber
\end{align}
where $\xi = r/r_0$. The constant is measured from the simulations (paper I).
The average specific angular momentum plotted against mass at varying radius is plotted in Fig. \ref{j_simu}
for the initial condition of four runs (solid curves) with varying levels of turbulent support, 
as well as the analytical solutions (dashed curves) with corresponding $v_\mathrm{rms}$.
In the mass range of our interest ($10^3-10^4$ M$_\odot$) , it roughly has the dependence $\overline{j}(r) \propto M(r)^{0.59}$. 

As discussed in Paper I, 
regardless of the angular momentum loss of sink particles (protostars),
the specific angular momentum of the gas component is more or less conserved at the proto-cluster scale during the collapse. 
We therefore do not consider its loss by angular momentum transport in our model.

The averaged specific angular momentum in the gaseous proto-cluster thus has the form:
\begin{align} 
j_\ast ~&\propto~ j_\mathrm{c} \left({M_\ast\over M_\mathrm{c}}\right)^{0.59} \propto~ \sigma_\mathrm{c} ~R_\mathrm{c} ~\alpha_{\ast,\mathrm{c}}^{-0.59}  
~\propto~ M_c^{(4-\gamma)/(6-2\gamma)} ~\alpha_{\ast,\mathrm{c}}^{-0.59}
\end{align}
As in the discussions for mass accretion, 
the quantities are estimated for the parent cloud and the mass ratio $\alpha_{\ast,\mathrm{c}}$ is used for the mapping.
The last approximation comes from the Larson relations (see Eqs. (\ref{Larson})).
Applying physical values gives:
\begin{align} 
j_\ast &= 6.7 \times 10^{19} \mathrm{m}^2 \mathrm{s}^{-1} {\sigma_\mathrm{rms} \over \sigma_\mathrm{vir}} \left({ \alpha_{\ast,c} M_\ast \over 10^4 \mathrm{ M}_\odot}\right)^{\gamma_j} \alpha_{\ast,\mathrm{c}}^{-0.59},
\label{eq_j}
\end{align}  
where $\gamma_j = 0.75$ or $0.72$ corresponding to $\gamma = 1$ or $0.7$.
This value is normalized with a $10^4$ M$_\odot$ cloud following Larson relations (which is close to virial equilibrium), 
and should be multiplied by a correction factor $\sigma_\mathrm{rms} / \sigma_\mathrm{vir}$ if the turbulence is not following such relations.

\subsection{Solving for the gaseous proto-cluster mass-size relation}

\begin{figure}[]
\centering
\includegraphics[width=0.5\textwidth]{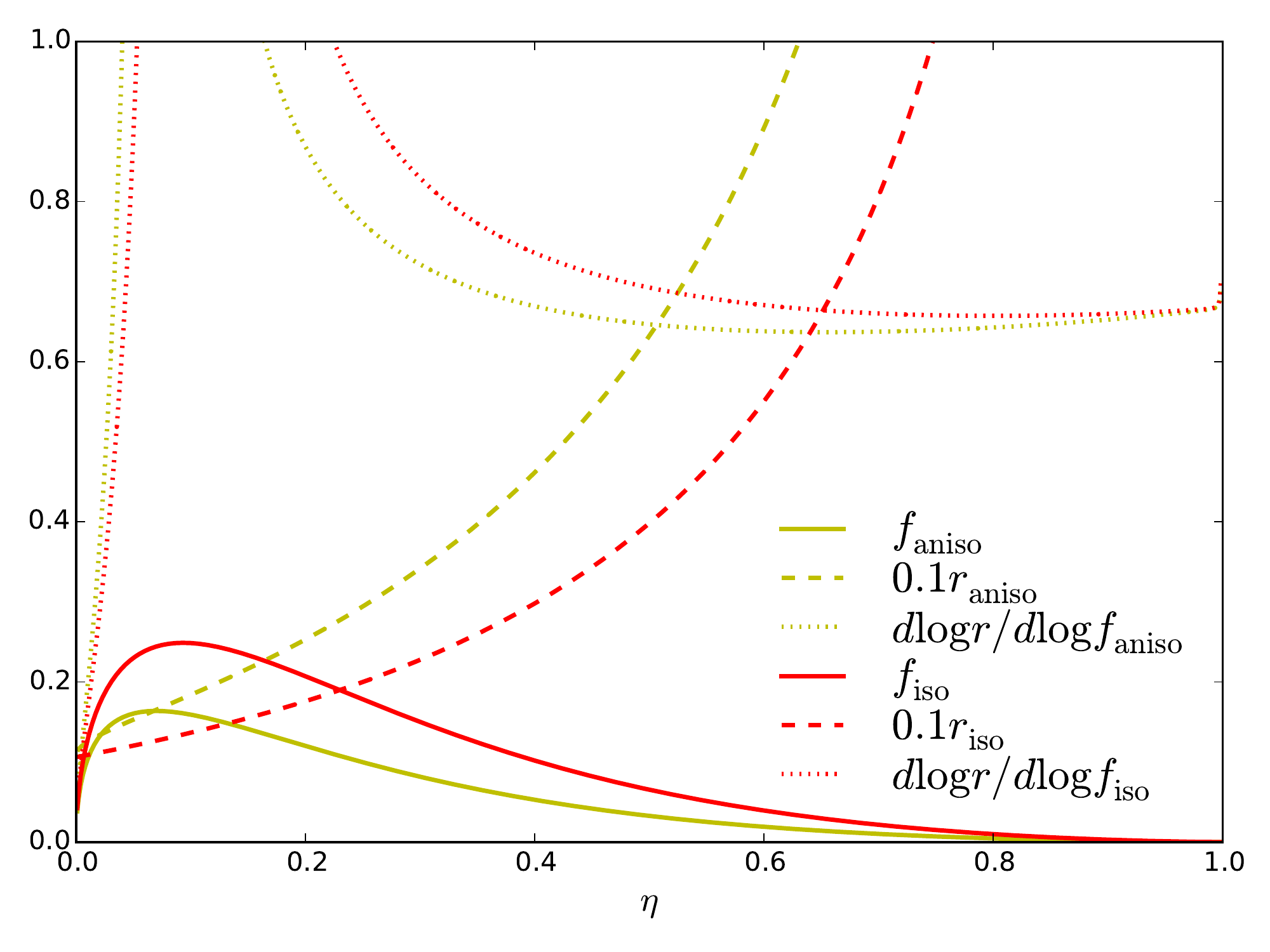}
\caption{The functions $f(\eta)$ (solid curves) and $r(\eta)$ (dashed curves, the value multiplied by 0.1 to be displayed in the same figure) for both anisotropic (yellow) and isotropic (red) assumptions of the turbulence. 
In dotted curves are the ratio between the changing rates of $r$ and $f$, plotted in absolute value.
}
\label{rf}
\end{figure}

Having obtained $\dot{M}$ and $j$ as functions of the gaseous proto-cluster mass $M$, 
Eqs. (\ref{eqnset}) 
could be solved to infer the three variables $R$, $\eta$, $\sigma^2$ for a given mass in equilibrium.
This is equivalent to solving for $\eta$ and $R$ from the equations:
\begin{align}
\label{feta}
f(\eta)=& {8 {5\over 2}^{3\over2} j^3 \dot{M} \over G^2 M^3} \\
=&\left\{ \begin{array}{lc}
          u_z^ {3 \over 2}(u_r-{1+\eta^{2\over 3}\over \eta^{2\over 3}}u_z)^{3 \over 2} (1+2\eta^{2\over 3}) / (\epsilon_\mathrm{acc} \eta u_g)& \mbox{anisotropic}  \\ 
          u_z^ {3 \over 2}(u_r-2 u_z)^{3 \over 2} 3 / (\epsilon_\mathrm{diss}\epsilon_\mathrm{acc} \eta u_g) & \mbox{isotropic} 
         \end{array}\right.\nonumber\\
\label{reta}
R =& {5j^2 \over 2GM} r(\eta) = \left\{ \begin{array}{lc}
           {5j^2 \over 2GM} \left(u_r-{1+\eta^{2\over 3}\over \eta^{2\over 3}}u_z\right)^{-1}  & \mbox{anisotropic}  \\ 
           {5j^2 \over 2GM} \left(u_r-2u_z\right)^{-1} & \mbox{isotropic} 
         \end{array}\right. 
\end{align}
The functions $f(\eta)$ and $r(\eta)$ are geometrical factors that depend on the aspect ratio $\eta$.
They are shown in Fig. \ref{rf} in solid and dashed curves  ($r(\eta)$ is multiplied by $0.1$ to display in the same figure)  respectively.
The anisotropic functions are plotted in yellow, and the isotropic ones in red.
We also plot in dotted curves $d\log{r}/d\log{f}$, the ratio between the relative growth rate of $r$ and $f$,
to highlight how the gaseous proto-cluster radius solution depends on the model parameters, and in turn on $f(\eta)$.
This is further discussed in \S \ref{dependence}.

The solution occurs at the intersection of $f(\eta)$ and a constant determined by $M$, $\dot{M}$ and $j$ as seen in Eq. (\ref{feta}).
Since $f(\eta)$ is first increasing and then decreasing in the interval $\eta \in [0,1]$, 
two solutions coexist when there are solutions.
It is the solution with larger value of $\eta$ that is physical, 
because it reduces to the spherical case when the angular momentum goes to zero.
The recovered $\eta$ value gives in turn the radius of the gaseous proto-cluster by Eq. (\ref{reta}).
In this solution, turbulent  energy dominates over rotational energy,
it is also confirmed from the simulation results (see paper I) that the rotational energy is small initially and accounts for $30-40\% $ of the total kinetic energy of the gaseous proto-cluster by the end of the simulation.

\begin{figure}[]
\centering
\includegraphics[trim=80 10 120 10,clip,width=0.5\textwidth]{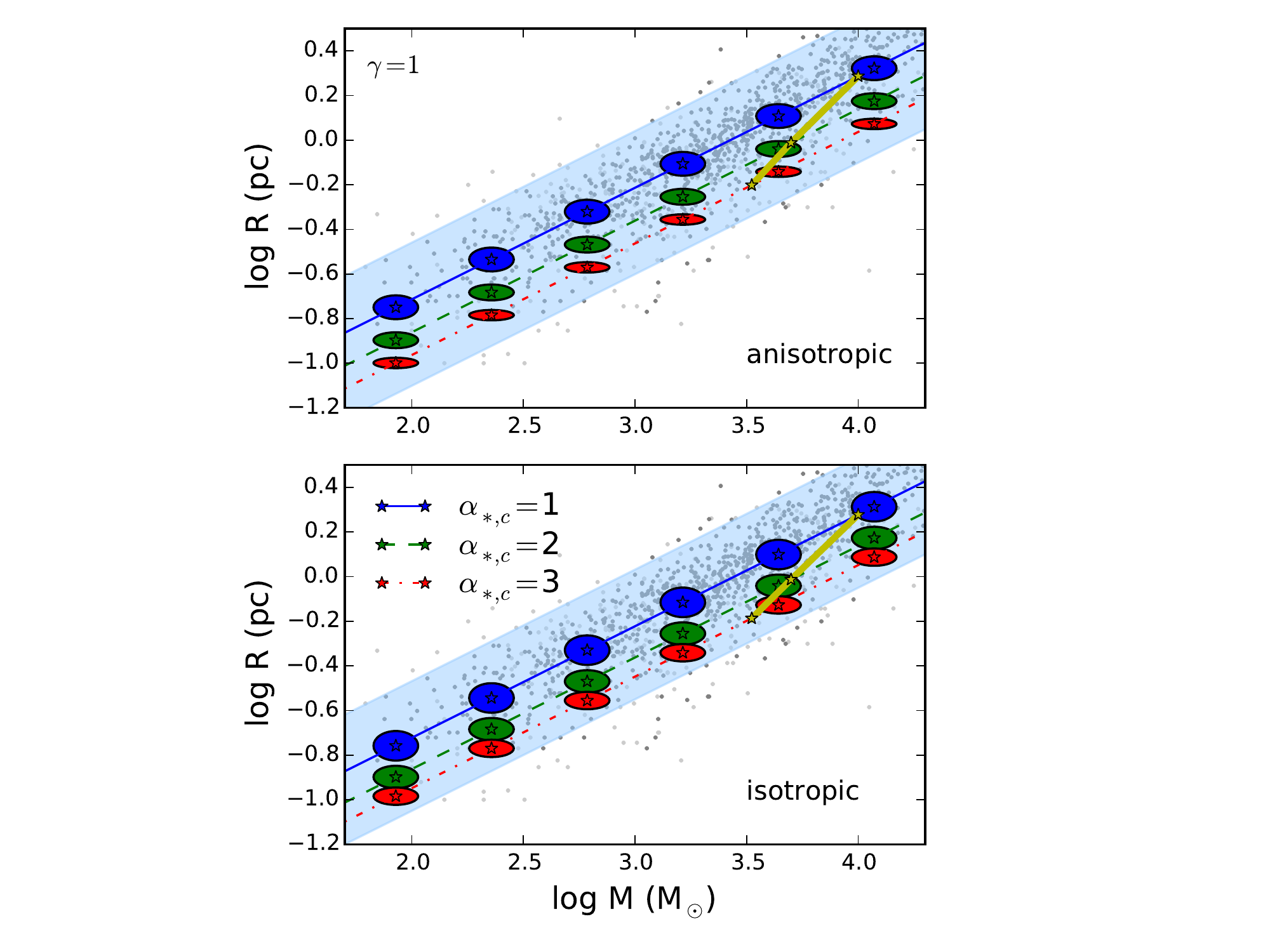}
\vspace{-1\baselineskip}
\caption{The mass-size relation of ellipsoidal clusters, 
shown with cloud-cluster mass ratio $\alpha_{\ast,\mathrm{c}} = 1, 2, 3$ plotted with blue solid, green dashed, red dot-dash lines respectively,
using anisotropic turbulence following Kolmogorov spectrum (upper panel)
and isotropic turbulence (lower panel). 
The turbulence efficiency $\varepsilon=0.5$ is used, 
and the shadowed region represents the range of solutions for $\varepsilon \in [0.2,0.7]$.
The elliptical patches represent the form of the clusters.
With the molecular cloud density-size relation $\rho \propto R^{-1}$, 
the clusters follow a $R \propto M^{0.5}$ trend for a given $\alpha_{\ast,\mathrm{c}}$. 
The gaseous proto-cluster corresponding to $10^4$ M$_\odot$ cloud is traced in yellow, 
which is indicative of an evolutionary sequence roughly following $R \propto M$. 
The radius of the gaseous proto-cluster does not depend too much on the underlying assumption of the turbulent energy distribution,
while the aspect ratio is smaller in the anisotropic case.
The dots are the observed star-forming clumps from \citet{Fall10} and \citet{Urquhart14}.
}
\label{elpsplot1}
\end{figure}

\begin{figure}[]
\centering
\includegraphics[trim=80 10 120 10,clip,width=0.5\textwidth]{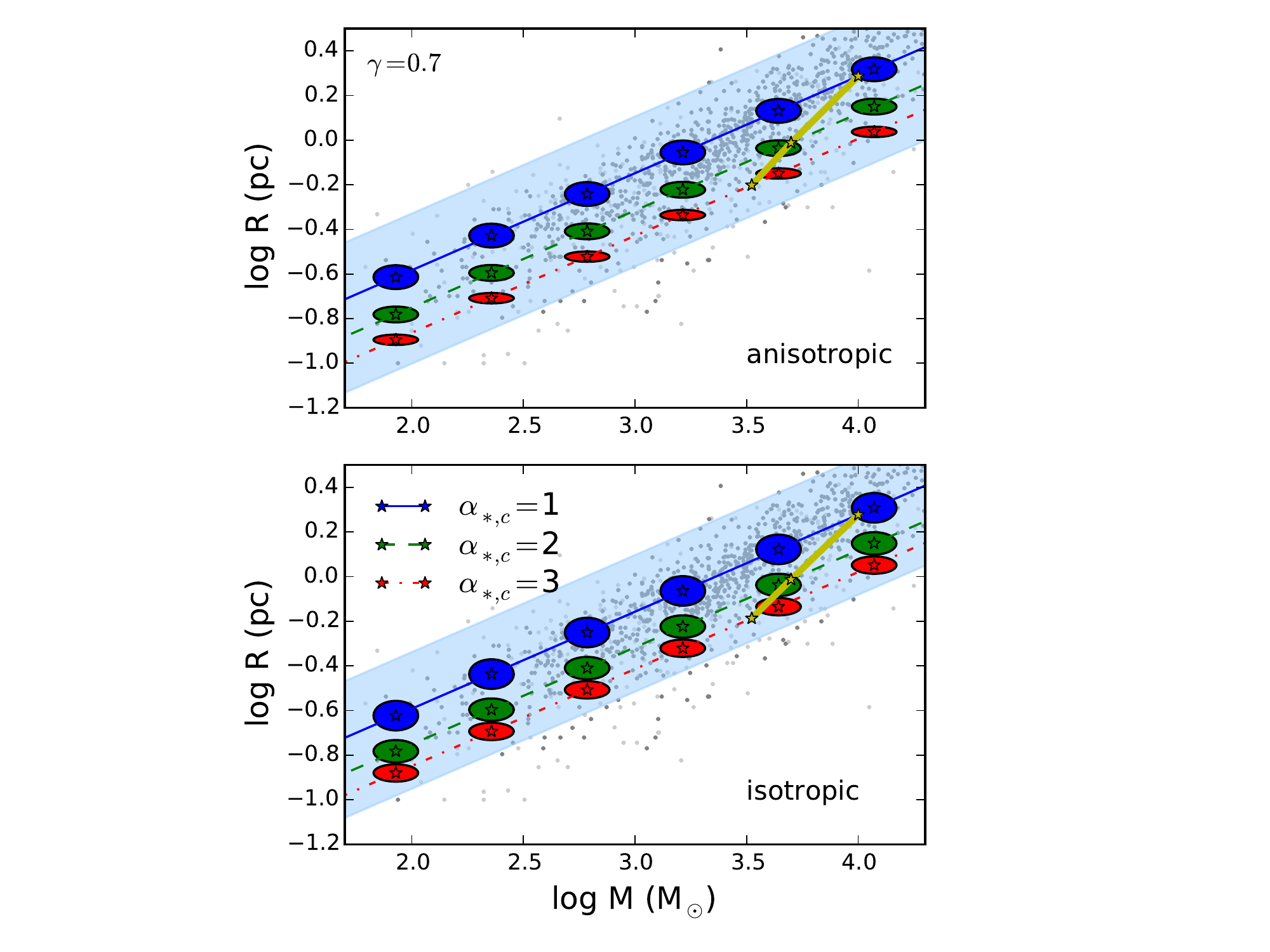}
\vspace{-1\baselineskip}
\caption{
The same plots as Fig. \ref{elpsplot1} for gaseous proto-clusters inside parent clouds following $\rho \propto R^{-0.7}$ relation. 
The mass-size relation follows $R \propto M^{0.42 - 0.44}$. 
}
\label{elpsplot07}
\end{figure}

The gaseous proto-cluster mass-size relation is governed by several parameters. 
We first apply some canonical values to illustrate the solution properties before confronting with simulation results.
Three cloud-cluster mass ratios $\alpha_{\ast,\mathrm{c}} = 1, 2, 3$ are used for virialized molecular clouds, 
i.e. $\sigma_\mathrm{rms} / \sigma_\mathrm{vir} =1$, in Eqs. (\ref{eq_mdot}, \ref{eq_j}).
The canonical turbulence driving efficiency $\varepsilon = 0.5$ is used 
(see Eqs. (\ref{eqn_iso}, \ref{eqnset}, \ref{feta}), 
$\varepsilon=\epsilon_\mathrm{acc}$ and $\varepsilon=\epsilon_\mathrm{diss}\epsilon_\mathrm{acc}$ respectively). 
The resulting mass-size relation is shown in Figs. \ref{elpsplot1} and \ref{elpsplot07} for two $\gamma$ values.
and the range of solutions with $\varepsilon \in [0.2,0.7]$ is represented by the shadowed region
to illustrate the dependence of the radius on $\varepsilon$.
The radius of the cluster decreases with increasing $\varepsilon$.
Nonetheless, the dispersion of observations is compatible with a large range of the model parameters, 
guaranteeing the robustness of our model prediction regardless of some poorly controlled factors.
Note that solutions might not exist for large $\varepsilon$, large $\alpha_{\ast,\mathrm{c}}$, 
or highly super-virial clouds. 

This is reasonable since large $\alpha_{\ast,\mathrm{c}}$ is interpreted as the early stage of gaseous proto-cluster formation, 
a high turbulence driving need a larger cluster-mass to keep itself bound, 
and a originally more turbulent cloud feeds the proto-cluster with more angular momentum and kinetic energy. 

Our two dimensional virial model yields a $R \propto M^{0.5}$ relation for $\gamma = 1$ 
and $R \propto M^{0.42-0.44}$ for $\gamma = 0.7$ depending on the parameters used. 
These power-law exponents are compatible with those from \citet{Fall10} (0.38) and \citet{Urquhart14} (0.50). 
We over-plot the model with their star-forming clumps. 
The trend is very closely reproduced, however there exist a slight shift, which could depend on the observation sensitivity, the definition of radius, and the fact that we use a uniform density model. 
The underlying assumptions of turbulent energy distribution in anisotropic configuration does not change too much the resulting gaseous proto-cluster size,
and its effect is rather pronounced in 
the aspect ratio of the gaseous proto-cluster.
The mass-size relations that we obtained is thus robust despite our ignorance of the properties of the turbulence.
Note that the curves do not represent evolutionary sequences.
After a gaseous proto-cluster reaches equilibrium, 
it evolves towards sequences with smaller $\alpha_{\ast,\mathrm{c}}$ presuming that the parent cloud is not accreting mass much faster than the gaseous proto-cluster.
We trace in yellow the gaseous proto-cluster formed inside a $10^4$ M$_\odot$ cloud in Figs. \ref{elpsplot1} and \ref{elpsplot07}, 
which is indicative as an evolutionary sequence following $R \propto M$. 
A gaseous proto-cluster in equilibrium could not exist for too small mass, and its maximum mass is limited by the parent cloud. 
Therefore the clumps should not migrate too much on this equilibrium sequence during their life span, 
and the mass-size relation observed for star-forming clumps is rather a result of gaseous proto-clusters forming from a range of different gas reservoir than an evolutionary track.

\section{Discussion and comparison}
Having obtained the two-dimensional virial theorem for the rotating and accreting gaseous proto-cluster,
we discuss in this section how the parameters affect the model prediction.
Most importantly, we yield a comparison with the gaseous proto-clusters formed in simulations (paper I).
Care must be taken in this comparison when adapting a pure gas model to simulations with sink particle prescriptions.

\subsection{Dependance of the equilibrium state on gaseous proto-cluster properties}
\label{dependence}
As described by Eq. (\ref{feta}), 
the equilibrium solution is determined by intersection of the function $f(\eta)$ with a constant proportional to $j^3 \dot{M} M^{-3}$, 
so is the existence of solutions dependent on these parameters.
First we discuss how the parameters affect the form of the cluster. 
The gaseous proto-cluster parameters are scaled self-similarly in Figs. \ref{elpsplot1} and \ref{elpsplot07}, 
while there would be deviations if they do not follow exactly the same relations. 
The solution in the regime where $f(\eta)$ is monotonically decreasing is considered. 
We can see from Fig. \ref{rf} that
the gaseous proto-cluster is flatter and more extended if rotational support is important. 
Higher accretion rate also results in flatter form probably because shorter crossing time is required to dissipate larger energy input. 
In terms of the mass, when a cluster is more massive, it tends to be more spherical. 
The yellow line in Figs. \ref{elpsplot1} and \ref{elpsplot07} is roughly indicating a track where the mass is increasing while accretion rate stays almost the same and the specific angular momentum grows not as rapidly as the mass (eq. (\ref{eq_j})). 
In this case the aspect ratio increases as expected from Eq. (\ref{feta}).

On the other hand, equilibrium cannot exist if the angular momentum is too high.
This could partly due to the simplifications of angular momentum conservation applied to the model.
If the angular momentum of the system becomes high enough,
its loss by angular momentum transport should no longer be neglected.
The dissipative loss should be considered to give a lower angular momentum,
therefore extending the range of equilibrium existence.
High mass accretion rate could result in high velocity dispersion which could no longer be balanced by gravity.
In this case, the ram pressure also becomes important and should not be neglected.
This will result in a system which is predominantly equilibrated by the supporting turbulence and the confining surface ram pressure.
However, this is beyond the scope of our discussion since the gaseous proto-clusters do not have such high accretion rate 
and its self-gravity dominates over the ram pressure. 
As for the mass of the gaseous proto-cluster,
no equilibrium solution could exist for systems of too low mass (for a given cloud mass),
which is obvious since this corresponds to the early stage of gaseous proto-cluster formation and the self-gravity is not yet important enough to confine the system.

The turbulence driving efficiency by accretion $\epsilon_\mathrm{acc}$ is a poorly understood factor.
Some energy is radiated away as heat at the accretion shock.
A too high efficiency results in non-existence of equilibrium, 
while this problem is less constrained in the isotropic case where we have an extra factor of $\epsilon_\mathrm{diss}<1$
and $f_\mathrm{iso} > f_\mathrm{aniso}$.
On the other hand, too low efficiency would give too extended gaseous proto-clusters.
This could be explained by Fig. \ref{rf}, in which $r(\eta)$ increases while $f(\eta)$ deceases with $\eta$.
In our parameter domain of clouds following Larson relations, 
the isotropic case can tolerate $\varepsilon=\epsilon_\mathrm{diss}\epsilon_\mathrm{acc}$ up to 1 while it is by definition less than 1, 
and the anisotropic case, on the other hand, has $\varepsilon=\epsilon_\mathrm{acc}$ limited to about 0.7 for larger $\alpha_{\ast,\mathrm{c}}$, say, 3. 
This means that if the turbulence-driving efficiency is indeed high, 
the turbulence inside the gaseous proto-cluster would prevent equilibrium and cause expansion at the beginning of its formation, and the quasi-static equilibrium could only be reached after enough mass is accreted. 
As shown in Fig. \ref{rf}, 
the radius ls less dependent of $\varepsilon$ at larger driving efficiency and a lower limit is well defined, 
while it grows very fast towards large values at small  $\varepsilon$. 
The model ceases to be valid when the gaseous proto-cluster radius comes close to that of the cloud, 
or equivalently, when $\varepsilon \lesssim 0.1$. 
This should correspond to the condition in which a gaseous proto-cluster has not yet formed inside the cloud. 
The infall is small compared to turbulence, and the not-well-oriented flow is dominated by dissipation and thus has low driving-efficiency. 
Our model accounts for the gaseous proto-cluster which has reached quasi-static equilibrium, 
that is, a centrally concentrated mass is readily marked inside the molecular cloud. 
In this regime, the model is robust as long as the turbulence-driving efficiency is not too small ($>0.1$). 
Nonetheless, more care should be taken is we wish to understand the formation stage of the gaseous proto-cluster. 
One interesting thing to be noted is that at larger $\eta$, $d\log r /d \log f$ is close to $-2/3$,
giving in turn $r \propto f^{-2/3}$ and thus from Eq. (\ref{reta}) we derive $R \propto M \dot{M}^{-2/3}$ independent of the angular momentum.
Using Eq. (\ref{mdot_m}), we obtain $R \propto M^{0.5}$ or $R \propto M^{0.43}$ for $\gamma = 1$ or $0.7$.
This means that if we simply consider a turbulent spherical model without rotation ($j \rightarrow 0$),
similar gaseous proto-cluster mass-size relation would be concluded and this goes back to the \citet{Hennebelle12} result while the general rotating motion and flattened geometry would be missed.

We discuss two possibilities of turbulent energy distribution in a non-spherically-symmetric system: 
anisotropy and isotropy.
In the isotropic case, $f(\eta)$ is larger than that in the anisotropic case,
meaning that the equilibrium solution exists for a larger range of model parameters.
Nonetheless, the recovered mass-size relation are not significantly different,
guaranteeing that we are rather robust to the assumptions.
Last but not least, it is important to keep in mind that the analytical study is based on uniform density assumption of the  gaseous proto-cluster,
each term in the equations may differ from the actual value by a factor of order unity in consequence.
However, our conclusions should stay valid qualitatively, 
and the energy analysis of simulated proto-clusters (see Paper I) indeed coincides with this picture of two-dimensional global virial equilibrium.

\subsection{Adapting the model parameters to simulations}

So far the analytical model includes only the gaseous component,
while in the simulations we performed (see paper I) sink particles are formed to follow the dense regions.
The analytical model should be adapted to yield a better comparison with simulations,
we therefore use the modified equation set for the 2D virial equilibrium:
\begin{subequations}
\begin{align} 
{5\over 2} M_\mathrm{g} \left({j \over R}\right)^2 + M_\mathrm{g} s_r(\eta) \sigma^2 &= {GM_\ast M_\mathrm{g}\over R} u_r(\eta) \\
M_\mathrm{g} s_z(\eta) \sigma^2 &= {GM_\ast M_\mathrm{g}  \over R} u_z(\eta) \\
M_\mathrm{g}d(\eta) { \sigma^3 \over  4 R  } 
&=  \epsilon_\mathrm{acc} {G(\dot{M}_\ast M_\mathrm{g}+M_\ast \dot{M}_\mathrm{g} )  \over R} u_g(\eta) ,
\label{vir2d} 
\end{align}
\end{subequations}
where $M_\mathrm{g}$ is the gas mass and $M_\ast$ is the cluster mass, sink mass included.
Note that the underlying assumption of this formalism is that both gas and sinks are uniformly distributed,
thus we are allowed to integrate over mass $M_\mathrm{g}$ the gravitational force created by mass $M_\ast$.
As seen in the simulations that the gas mass inside the cluster stays relatively constant and most of the mass is accreted onto the sinks,
we make the simplification $\dot{M}_\mathrm{g} = 0$ and the variable $M_\mathrm{g}$ could be eliminated from all the equations.
The equation set remains almost unchanged with respect to the pure gas model (Eqs. (\ref{eqnset})),
except a factor 2 in the energy released from accretion.
Feedback mechanisms are not taken into account in the simulations.
In more realistic runs where accretion is partly prevented by feedback,
we should have something between this setup and the pure gas description.
This adaptation turns out not to influence much the model predictions, 
which should be therefore robust during the early proto-cluster evolution before stellar feedbacks (not treated in this study) significantly modify the cluster environment.
Note that the mass intervening in the model, 
also from which the accretion rate and the angular momentum is evaluated,
is the total cluster mass instead of the gas mass.

\subsection{The cluster mass-size relation compared to simulations}
\label{sec_m_s}
\begin{figure}[]
\centering
\begin{subfigure}{0.5\textwidth}
\includegraphics[trim=80 10 120 10,clip,width=\textwidth]{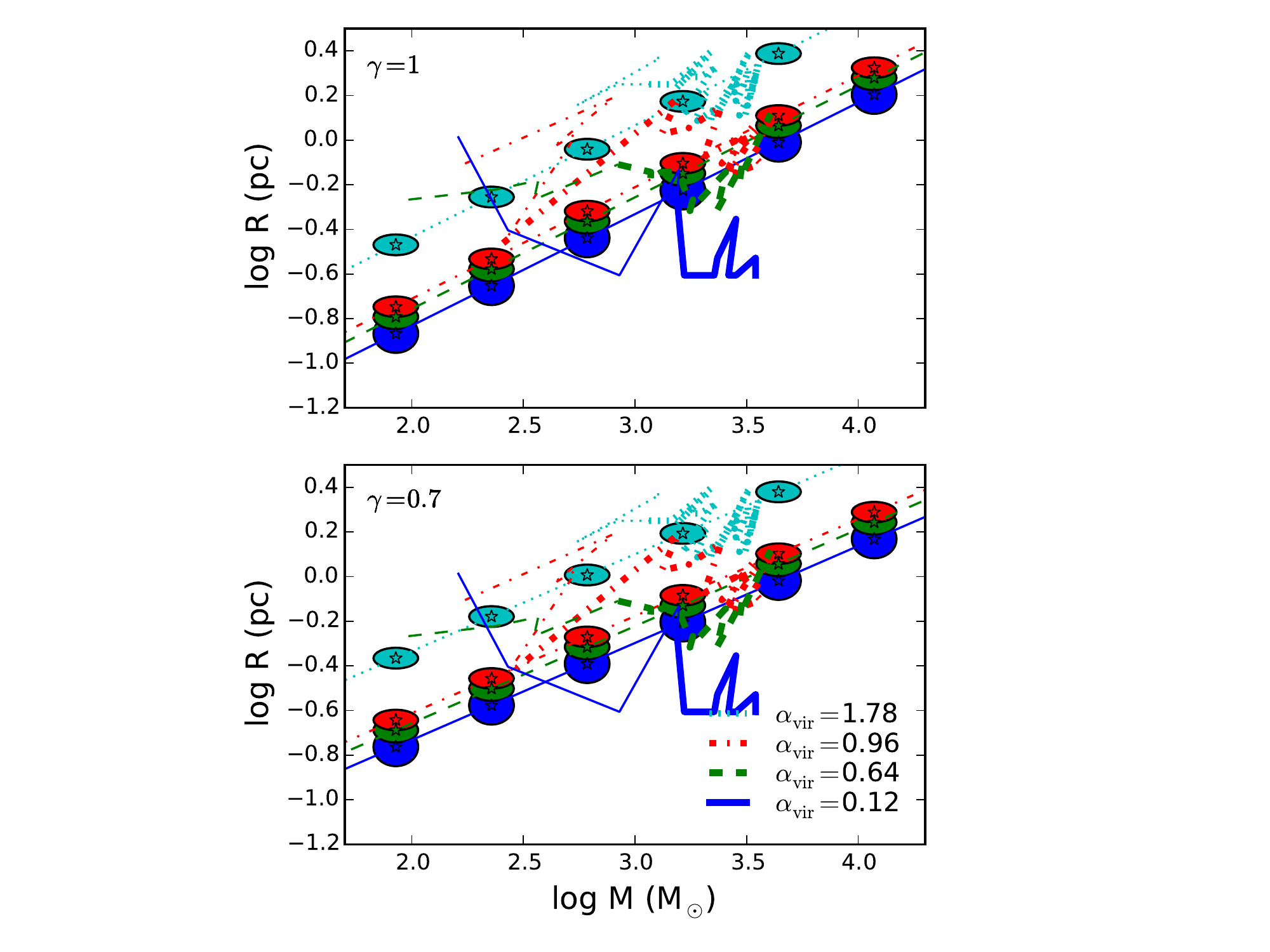}
\end{subfigure}
\caption{The mass-size relation of ellipsoidal clusters, the total mass inside the ellipsoidal region defined with gas kinematics over-plotted with model of corresponding levels of turbulence with $\gamma=1$  (upper panel) and $0.7$ (lower panel). 
The velocity dispersion increases from the bottom curve to the top one. 
Simulation results at time before 2 Myr are plotted with thin lines. 
The model is in good agreement with simulation only after the proto-cluster gains enough mass, 
possibly implying that the time dependent terms and ram pressure should not be neglected at early stage.
}
\label{virialsimu}
\end{figure}

In paper I, 
we evaluate the proto-cluster mass and size using gas kinematics and sink particle distributions.
The gaseous proto-cluster mass-size relations determined with gas kinematics are over-plotted with models with corresponding levels of turbulence for four simulations in Fig. \ref{virialsimu}. 
We do not discuss the sink cluster since our model considers only the gas component. 
Readers are invited to refer to paper I for a comparison between the gas and sink clusters.  
The level of turbulent support is represented by the  viral parameter $\alpha_\mathrm{vir} = 2E_\mathrm{kin}/E_\mathrm{grav}$.
The angular momentum in the model is multiplied by the ratio between the level of turbulence in each of the runs and the virialized value, 
the mass accretion rate is also calculated accordingly.
We show models with the values $1$ and $0.7$ for the molecular cloud Larson relation $\rho \propto R^{-\gamma}$.
Canonical values of cloud-cluster mass ratio $\alpha_{\ast, \mathrm{c}} = 3$ and $\varepsilon = 0.5$ are used along with isotropic turbulence.

The earlier time steps (before 2 Myr) in simulations are shown with thinner lines.
The results show that the proto-cluster could be identified at a relative early stage when the mass is still small,
and that as the gaseous proto-cluster accretes mass, it arrives on the sequence where virial equilibrium is reached.
The model does not explain well the gaseous proto-cluster at the early stage,
possibly because the time dependent terms and ram pressure are still relatively important and should not be neglected, while we use quasi-stationary assumptions.
Once the mass is large enough,
the simulations are in good agreement with the model. 
The case with $\alpha_\mathrm{vir} = 0.12$ is probably too strongly accreting due to the weak kinetic support, 
thus does not correspond well to the quasi-stationary model. 
Otherwise, though showing slightly different trends, 
the model with two $\gamma$ values are both compatible with the simulation results. 

\subsection{The IMF peak position}
\begin{figure}[]
\centering
\includegraphics[trim=65 10 65 10,clip,width=0.5\textwidth]{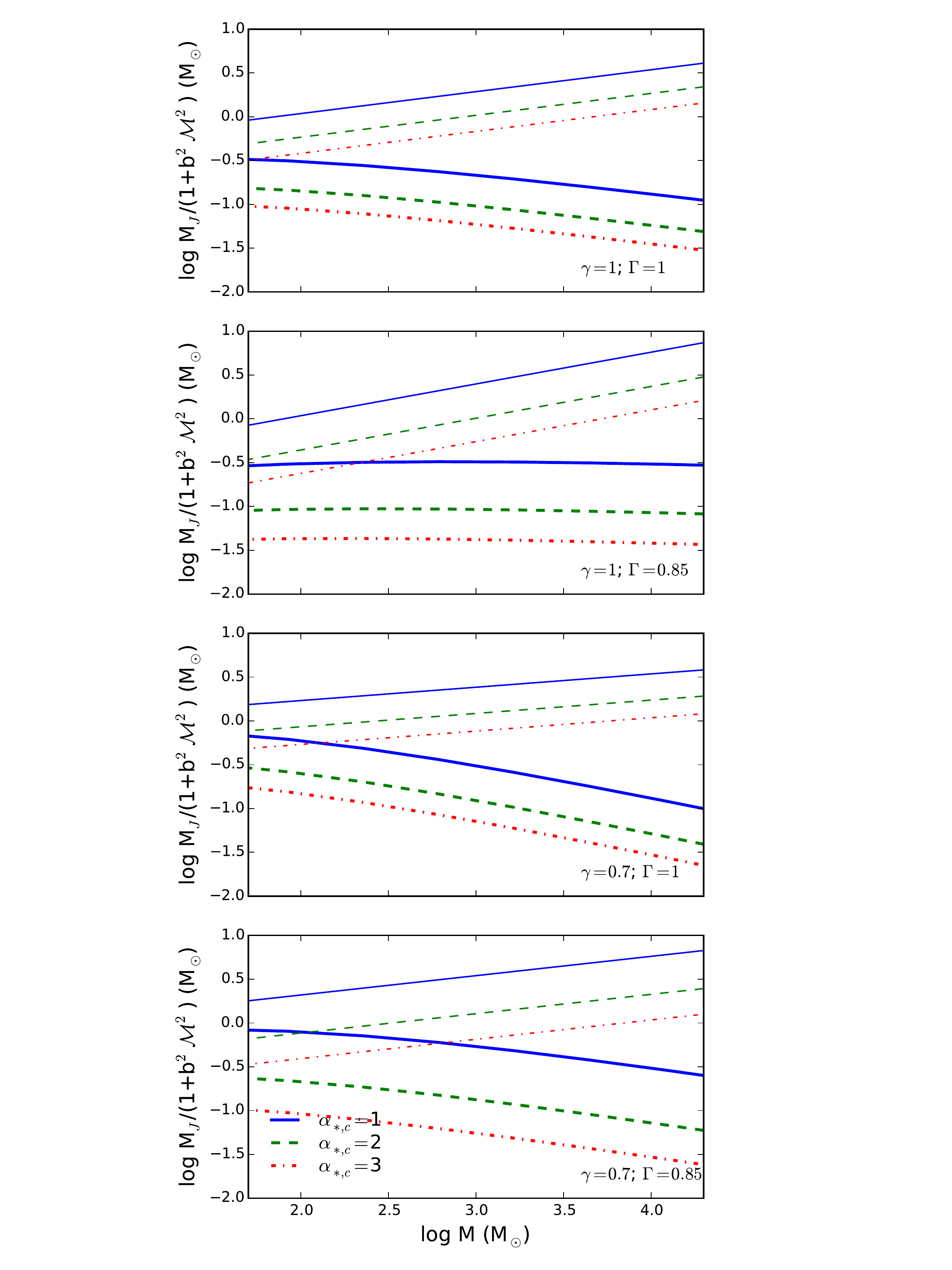}
\caption{The IMF peak prediction from proto-cluster model with isotropic turbulence for cloud-cluster mass ratio $\alpha_{\ast,\mathrm{c}} = 1, 2, 3$ plotted against proto-cluster mass.
The values $1$ and $0.7$ are used for the Larson relation $\rho \propto R^{-\gamma}$. 
As for the polytropic index $\Gamma$ of gas inside the proto-cluster, values $1$ (isothermal) and $0.85$ are considered.
The Jeans mass is plotted with thin curves, and the IMF peak mass is plotted with thick curves.
Color codings are same as that  in Fig. \ref{elpsplot1}.}
\label{Mpeak}
\end{figure}

The theoretical prediction of the CMF peak position \citep{HC13} depends on the Jeans mass and the Mach number of the star-forming gas.
We calculate for our model 
the Jeans mass $M_\mathrm{J} = \pi^{5/2} c_s^3 / 6 \sqrt{G^3 \rho}$ (thin curves) and the predicted CMF peak mass $M_\mathrm{peak} = M_\mathrm{J}/(1+b^2\mathcal{M}^2)$ (thick curves), 
which are shown in Fig. \ref{Mpeak}.  
A canonical value $b = 0.5$ is used \citep{Federrath10}. 
We present the result with two $\gamma$ values (see Eq. (\ref{Larson})). 
As for the sound speed, we consider either the isothermal case at 10 K, 
or gas following polytropic relation $P \propto \rho^{\Gamma}$, where $\Gamma = 0.85$ \citep{Hennebelle12}.

The average density decreases with gaseous proto-cluster mass, and the Jeans mass is thus increasing.
On the other hand, 
the turbulence increases with gaseous proto-cluster mass and gives a reasonable IMF peak prediction
assuming that CMF and IMF peaks coincide. 
In the mass range between $10^2$ and $10^4$ solar mass, 
where most clusters are observed, 
the predicted IMF peak value is around $0.1-0.3$ M$_\odot$ with less than 1 dex variation. 
The molecular clouds following $\rho \propto R^{-1}$ give a flatter relation. 
It is probably too simplifying to assume same temperature for all clusters, 
since higher mass clusters have lower density and thus slightly higher temparature. 
The theoretical peak position depends on the temperature as $T^{5/2}$, 
thus the variation could be further reduced if this effect is taken into account.
With the polytropic index $\Gamma=0.85$, the inferred peak mass is more uniform among clusters of different mass, 
suggesting that an isothermal approach might be too simplistic.
The observed IMF peak is about $0.3$ M$_\odot$ with some dispersions \citep[e.g.][]{Bastian10}, 
which is slightly higher than what we discover here particularly if we would take into account 
the apparent shift that is observed between the CMF and the IMF \citep[e.g.][]{andre2010}.
However, the magnetic field, which provides additional support against self-gravity, 
is not considered, 
and can shift the peak towards larger value when incorporated.
Furthermore, when stars start forming, this environment may be self-regulated due to stellar feedback.
Nonetheless, this simple prediction gives a general picture which is compatible with a universal IMF peak.

\section{Conclusions}

As it is known that a significant fraction of stars form in clusters, 
it is fundamental understanding how these latter form and what their physical characteristics are.
In this work we confronted observations and simulation results with a simple analytical model, 
and showed that before stars start forming, 
the molecular cloud gas is reprocessed and a gaseous proto-cluster environment in global energy equilibrium is established, 
which is much more favorable for star formation than the more diffuse large scale molecular clumps. 
This gaseous proto-cluster sets a more general condition for stellar cluster formation and to some extent decouples star formation from the large scale molecular clump. 
This is compatible with the idea that stellar clusters form in similar environments and that the IMF is regulated by more local conditions.

We developed an analytical model to account for the gaseous proto-cluster in virial equilibrium.
A two-dimensional model is derived for a system with rotation, turbulence, and accretion,
which predicts the radius, aspect ratio and velocity dispersion given the mass, angular momentum and accretion rate of the system.
The mass accretion rate is estimated using free-fall collapse of a molecular cloud in which resides the gaseous proto-cluster,
while adding a correction for varying level of turbulent support.
The angular momentum of the gaseous proto-cluster is estimated using residual turbulent vortices while assuming no loss by transport.
Its absolute value is scaled to coincide with that in the simulations.
Given these estimations,
we obtained an ellipsoidal virialized structure which is supported by rotation and supersonic turbulence against self-gravity.
We produced a mass-size relation for gaseous proto-clusters which is in coherence with observational \citep{Fall10, Urquhart14} and simulation  results (paper I).
The model dependence of the parameters was also discussed and we found our model to be quite robust in predicting the gaseous proto-cluster properties 
regardless of the turbulence nature, the turbulence driving efficiency, and the infalling flow pattern which are not well known. 

We conclude that the gaseous proto-clusters lie on an equilibrium sequence which is governed by the interaction of gravity and turbulence.
This yields a mass-size relation similar to the Larson's relation, 
while a gaseous proto-cluster is roughly 10 times more massive than a molecular cloud of the same size. 
Such resemblance is seen in the two relations since both are outcomes of turbulence and gravity interaction,  
while we emphasize that in the case of a gaseous proto-cluster, 
the accretion is relatively important in concentrating the mass and sustaining the turbulence, 
thus creating an environment different from that of the molecular cloud.
As most stars form in clusters, 
the gaseous proto-cluster gas properties should be indeed used for understanding star formation.
Using a simple estimate for the peak position of the core mass function, we show that, due to various compensations,
 it depends only weakly on the 
cluster mass, which suggests that the physical conditions of gaseous proto-clusters may be, at least in part, 
responsible for the apparent universality of the IMF.


\begin{acknowledgements}
This work was granted access to HPC
   resources of CINES under the allocation x2014047023 made by GENCI (Grand
   Equipement National de Calcul Intensif). 
   This research has received funding from the European Research Council under
   the European Community's Seventh Framework Programme (FP7/2007-2013 Grant
   Agreement no. 306483). 
   The authors thank the anonymous referee for the careful reading and useful suggestions. 
\end{acknowledgements}



\appendix
\section{Two dimensional virial integration}
\label{appen_vir}
The virial theorem is conventionally used for uniform density spheres.
Here we deduce a more general form for an oblate ellipsoidal cluster of uniform density. 
The gravitational potential inside a uniform density ellipsoid is given by \citep{Neutsch79}:
\begin{align} 
\phi(x,y,z) =& - G \rho \pi abc \int_0^\infty \frac{1- {x^2 \over a^2 + \lambda} - {y^2 \over b^2 + \lambda} - {z^2 \over c^2 + \lambda}}{\sqrt{(a^2 + \lambda)(b^2 + \lambda)(c^2 + \lambda)}} d \lambda  \\
=&  -G \rho \pi R^2H \int_0^\infty \frac{1- {r^2 \over R^2 + \lambda} -  {z^2 \over H^2 + \lambda}}{(R^2 + \lambda)\sqrt{H^2 + \lambda}} d \lambda,\nonumber 
\label{pot_elp} 
\end{align}
where the general form of an ellipsoid with semi-axes $a \geq b\geq c$ is applied to our oblate ellipsoid of semi-axes $R$ and $H$.
Cylindrical coordinate is used.
This allows to integrate the right hand side of the virial equation in $r$-direction and gives a term analogous to gravitational potential energy, 
while being the result of only two of the dimensions of the gravity.
Same is the case for $z$-direction potential.

Let us first consider the multiplying by the $\vec{r}$ vector in a plane:
\begin{align} 
\int \limits_V \rho d_t \vec{v}  \cdot \vec{r} dV=& \int \limits_V - \rho \nabla \phi \cdot \vec{r}dV 
\end{align}
The left hand side becomes:
{\allowdisplaybreaks\begin{align} 
&\int \limits_V \rho d_t \vec{v}  \cdot \vec{r} dV \\
= & \int \limits_V  \rho d_t(\vec{v}\cdot \vec{r})  dV - \int \limits_V  \rho \vec{v}\cdot \vec{v}_\mathrm{2d}  dV \nonumber\\
=& \int \limits_V \rho \partial_t (\vec{v}\cdot \vec{r}) dV +  \int \limits_V \rho \vec{v}\cdot \nabla (\vec{v}\cdot \vec{r}) dV  - \int \limits_V  \rho v_\mathrm{2d}^2  dV\nonumber\\
=&   \partial_t \int \limits_V \rho (\vec{v}\cdot \vec{r}) dV +  \int \limits_V \nabla\cdot  (\vec{v}\cdot \vec{r}\rho \vec{v}) dV  - \int \limits_V  \rho v_\mathrm{2d}^2 dV\nonumber\\
=&   \partial_t \int \limits_V \rho d_t({ \vec{r} \cdot \vec{r} \over 2}) dV +  \int \limits_S  \vec{v}\cdot \vec{r} \rho \vec{v}\cdot \vec{dS}  - \int \limits_V  \rho v_\mathrm{2d}^2 dV\nonumber\\
=&  \partial_t\int \limits_V \rho \partial_t( {r^2\over 2}) dV +  \partial_t\int \limits_V \rho \vec{v} \cdot \nabla ( {r^2\over 2}) dV  \nonumber\\
& \;  +  \int \limits_S  v_r r \rho \vec{v}\cdot \vec{dS}- \int \limits_V  \rho v_\mathrm{2d}^2 dV  \nonumber\\
=&  {1\over 2} \partial_t^2 \int \limits_V \rho r^2 dV + {1\over 2} \partial_t \int \limits_S \rho \ r^2 \vec{v} \cdot \vec{dS} \nonumber\\
& \;  +  \int \limits_S  v_r r \rho \vec{v}\cdot \vec{dS}  - \int \limits_V  \rho v_\mathrm{2d}^2 dV,\nonumber  
\end{align}}
where $u_{2d}$ is the velocity in the $x-y$ plane, and $u_r$ is the velocity in the $\vec{r}$ direction.
The right hand side becomes (while assuming uniform density):
{\allowdisplaybreaks\begin{align} 
& \int \limits_V - \rho \nabla \phi \cdot \vec{r}dV \\
 =& - \rho\iint    \partial_r \phi r2\pi r dr dz  \nonumber\\
 =& -  \rho\int   (\phi r 2\pi r) |_{r=0}^{r_s(z)} dz + \rho\iint   2\phi  2\pi r dr dz\nonumber\\
 =& 2\rho \int   (\int \phi  2\pi r dr -  \phi(r_s) \pi r_s^2 ) dz\nonumber\\
 =& 2\rho \int \limits_V (\phi(r,z) - \phi(r_s,z)) dV \nonumber \\
 =& -2G\rho^2 \pi R^2H  \int \limits_V \int_0^\infty \frac{1- {r^2 \over R^2 + \lambda} -  {z^2 \over H^2 + \lambda}}{(R^2 + \lambda)\sqrt{H^2 + \lambda}} -\frac{1- {r_s^2 \over R^2 + \lambda} -  {z^2 \over H^2 + \lambda}}{(R^2 + \lambda)\sqrt{H^2 + \lambda}} d \lambda dV \nonumber \\
=& -2G \rho^2 \pi R^2H  \int \limits_V \int_0^\infty R^2 \frac{1- {r^2 \over R^2} -  {z^2 \over H^2 }}{(R^2 + \lambda)^2\sqrt{H^2 + \lambda}} d \lambda dV\nonumber \\
=&   -2G \rho^2 \pi R^2H { R^2 \over (R^2 -H^2)^{3\over 2}} \left[\cos^{-1}{(\eta)} - \eta\sqrt{1-\eta^2}\right] \times \nonumber\\
& \; \int (1- {r^2 \over R^2} -  {z^2 \over H^2} ) dV \nonumber \\
=& -{3\over 5} {GM^2\over R}   {1 \over (1-\eta^2)^{3\over 2}} \left[\cos^{-1}{(\eta)} - \eta\sqrt{1-\eta^2}\right] , \nonumber
\label{mom_eq} 
\end{align}}
where $\eta = {H\over R}$ represents the aspect ratio of the ellipsoid, and $r_s(z)$ is the $r$ value corresponding to a given $z$ on the surface of the ellipsoid.
The last integration could be done with simple geometrical argument by integrating inside a sphere and then re-scaling with the aspect ratio:
{\allowdisplaybreaks\begin{align} 
\int_\mathrm{ellp} (1- {r^2 \over R^2} -  {z^2 \over H^2} ) dV &= \\
\eta \int_\mathrm{sph} (1- {r^2 \over R^2}) dV &= \eta {2\over 5} V_\mathrm{sph} = {2\over 5} V_\mathrm{ellp}.\nonumber
\end{align}}
Note that $r$ in the ellipsoid is the distance to the semi-minor axis and that in the sphere is the distance to the center.

The virial equation in the $\vec{z}$ direction gives:
\begin{align} 
\int \limits_V \rho d_t \vec{v}  \cdot \vec{z} dV=& \int \limits_V - \rho \nabla \phi \cdot \vec{z}dV 
\end{align}
Similarly, the left hand side becomes
\begin{align} 
&\int \limits_V \rho d_t \vec{v}  \cdot \vec{z} dV \\
=&  {1\over 2} \partial_t^2 \int \limits_V \rho z^2 dV + {1\over 2} \partial_t \int \limits_S \rho \ z^2 \vec{v} \cdot \vec{dS} +  \int \limits_S  u_z z \rho \vec{v} \cdot \vec{dS}  - \int \limits_V  \rho v_\mathrm{1d}^2 dV,\nonumber  
\end{align}
where $u_{1d}$ is the velocity in the $\vec{z}$ direction, and is equivalent to  $u_z$;
and the right hand side becomes
{\allowdisplaybreaks\begin{align} 
& \int \limits_V - \rho \nabla \phi \cdot \vec{z}dV  \\
 =& - \rho\iint    \partial_z \phi z2\pi r dz dr  \nonumber\\
 =& -  \rho\int  (\phi z 2\pi r) |_{z=-z_s(r)}^{z_s(r)} dr + \rho\iint   \phi  2\pi r dz dr\nonumber\\
 =& \rho \int   (\int \phi  dz 2\pi r -  \phi(z_s) 2z_s 2 \pi r ) dr\nonumber\\
 =& \rho \int \limits_V (\phi(r,z) - \phi(r,z_s)) dV \nonumber \\
 =& -G\rho^2 \pi R^2H  \int \limits_V \int_0^\infty \frac{1- {r^2 \over R^2 + \lambda} -  {z^2 \over H^2 + \lambda}}{(R^2 + \lambda)\sqrt{H^2 + \lambda}} -\frac{1- {r^2 \over R^2 + \lambda} -  {z_s^2 \over H^2 + \lambda}}{(R^2 + \lambda)\sqrt{H^2 + \lambda}} d \lambda dV \nonumber \\
=& -G \rho^2 \pi R^2H  \int \limits_V \int_0^\infty H^2 \frac{1- {r^2 \over R^2} -  {z^2 \over H^2 }}{(R^2 + \lambda)\sqrt{H^2 + \lambda}^3} d \lambda dV\nonumber\\
=& -G \rho^2 \pi R^2H  \left\{{2H\over R^2}-{2H^2 \over (R^2 -H^2)^{3\over 2}} \left[\cos^{-1}{(\eta)} - \eta\sqrt{1-\eta^2}\right]\right\} \times \nonumber \\
& \; \int (1- {r^2 \over R^2} -  {z^2 \over H^2} ) dV \nonumber \\
=& -{3\over 5} {GM^2\over R}    \left[ {\eta \over 1-\eta^2 } -{\eta^2 \cos^{-1}{(\eta)} \over (1-\eta^2)^{3\over 2}} \right] \nonumber ,
\end{align}}
where $z_s(r)$ is the $z$ value corresponding to a given $r$ on the surface of the ellipsoid.
We therefore decompose the virial theorem into two equations for two directions which are balanced respectively.

\section{Free-fall time evaluated at gaseous proto-cluster radius}
\label{appen_ff}
The free-fall time typically used represents the formation time of an infinitely small object by free collapse without any support.
Here we make a correction for free-falling collapse onto the gaseous proto-cluster,
which has comparable size to its parent cloud.
The mass in free fall follows the equation for any radius $r$ within the cloud region:
\begin{align}
\ddot{r} = -{GM(r) \over r^2},
\end{align}
where $G$ and $M(r)$ are the gravitational constant and the mass contained inside radius $r$.
Multiplying by $\dot{r}$ on both sides gives
{\allowdisplaybreaks\begin{align}
{1\over 2} d_t \dot{r}^2 &= d_t{GM(r) \over r}\\
\left( {dr \over dt } \right)^2 &= {2GM \over r} - {2GM \over r_i}
\end{align}
\begin{align}
t_\mathrm{ff}(r_i) &= \int_0^{t_\mathrm{ff}} dt =\int_{r_i}^{r_f} -\left({2GM \over r} - {2GM \over r_i}\right)^{-{1\over2}} dr \\
&=  \sqrt{{3\over 8\pi G \overline{\rho}(r_i)}} \int_1^{r_f\over r_i} -\sqrt{{\xi \over 1-\xi}} d\xi \nonumber\\
&= \sqrt{{3\over 8\pi G \overline{\rho}(r_i)}} \left[\cos^{-1}{\sqrt{{r_f\over r_i}}}+\sqrt{{r_f\over r_i}\left(1-{r_f\over r_i}\right)} \nonumber\right].
\end{align}}
We thus have the time that the mass originally situated at radius $r_i$ takes to arrive at $r_f$,
given $\overline{\rho}(r_i)$, the initial averaged density inside $r_i$.
At $r_f=0$, this converges to the conventional free-fall time $\sqrt{3\pi \over 32 G \overline{\rho}}$.

\section{The decomposed ram pressure}
\label{appen_pram}
When integrating the virial equations,
a term analogous to the ram pressure in the spherical model \citep{Hennebelle12} appears,
while the ellipsoidal geometry renders its interpretation less obvious since we are ignorant of the mass infall pattern.
As long as we have $\dot{M} = -\int \limits_S \rho_\mathrm{inf} \vec{v}_\mathrm{inf} \cdot \vec{dS}$,
the integrals could be expressed in the following form by assuming that the gas reaches free-fall velocity upon accretion:
\begin{align} 
\int \limits_S  v_r r \rho \vec{v}\cdot \vec{dS} &= \dot{M}\sqrt{2GMR}p_r(\eta) \\
\int \limits_S  v_z z \rho \vec{v}\cdot \vec{dS} &= \dot{M}\sqrt{2GMR}p_z(\eta),
\end{align}
where $p_r(\eta)$ and $p_z(\eta)$ are dimensionless factors as functions of the ellipsoid aspect ratio.

Two extremes cases could be easily examined:
the accretion coming entirely along the edge or the pole of the ellipsoid.
In the first case we have $p_r=1$ and $p_z=0$, and in the second $p_r=0$ and $p_z=\sqrt{\eta}$.
A more sophisticated estimation is made by assuming that $\rho_\mathrm{inf} \vec{v}_\mathrm{inf}$ is constant and in radial direction on the whole ellipsoid surface  
and that $\vec{v}_\mathrm{inf}$ has constant value $\sqrt{2GM/R}$.
This allows to integrate the mass accretion rate therefore giving $p_r(\eta)$ and $p_z(\eta)$:
\begin{align} 
\dot{M} = -\rho_\mathrm{inf} \vec{v}_\mathrm{inf} 4 \pi R^2 \frac{\eta}{2\sqrt{1-\eta^2}} \left[{\pi \over 2} - \arctan{\left({2\eta^2-1 \over 2\eta\sqrt{1-\eta^2}}\right)} \right] \\
\int \limits_S  v_r r \rho \vec{v}\cdot \vec{dS} = 4 \pi \rho_\mathrm{inf} v_\mathrm{inf}^2 R^3{\eta \over 1-\eta^2} \left[ 1-{ \eta^2 \mathrm{arcsech}(\eta) \over \sqrt{1-\eta^2}} \right] \\
\int \limits_S  v_z z \rho \vec{v}\cdot \vec{dS} = 4 \pi \rho_\mathrm{inf} v_\mathrm{inf}^2 R^3{\eta^3 \over 1-\eta^2} \left[ {\mathrm{arcsech}(\eta) \over \sqrt{1-\eta^2}}-1 \right] 
\end{align}
They are presented in Fig. \ref{uruzug} with other coefficients.
\begin{figure}[]
\centering
\begin{subfigure}{.5\textwidth}
\includegraphics[width=\textwidth]{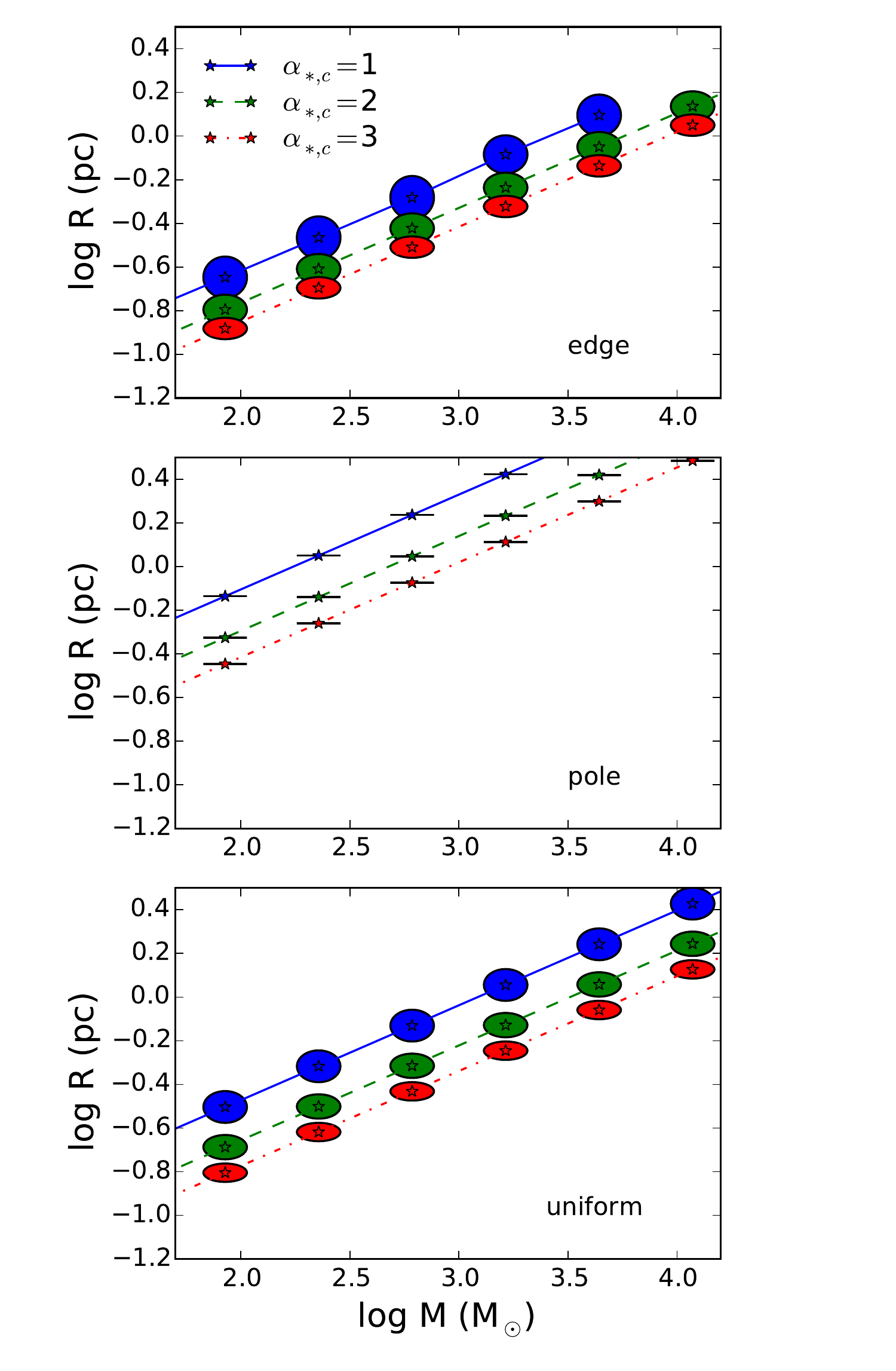} 
\end{subfigure}
\caption{
The mass-size relation of ellipsoidal clusters with mass infall along the edge (top),
with mass infall along the pole (middle), 
and with a more realistic and uniform infall estimation (bottom),
all other parameters are the same as that of the anisotropic model in Fig. \ref{elpsplot07}.
The clusters become more spherical when accretion comes from the sides,
while being completely flattened when flows come from the pole directions.
The results are almost unaltered compared to that without ram pressure in the case where the accretion comes from all directions,
only that the range of solution is slightly increased due to ram pressure confinement.
In all the three case,
the mass-size relation does not differ too much from one another.}
\label{pramplots}
\end{figure}

The more complete equation set becomes:
\begin{subequations}
\begin{align} 
{5 \over 2} \left({j \over R}\right)^2 + s_(\eta) \sigma^2 &= {GM \over R} u_r(\eta) + \dot{M}\sqrt{{2GR\over M}}p_r(\eta)\\
s_z(\eta) \sigma^2 &= {GM \over R} u_z(\eta) + \dot{M}\sqrt{{2GR\over M}}p_z(\eta) \\
d(\eta) { \sigma^3 \over  4 R  } 
&=  \epsilon_\mathrm{acc}{2G\dot{M} \over R} u_g(\eta),
\end{align}
\end{subequations}
which we evaluate for the three infall patterns. 
We show the mass-size relations for the three cases in Fig. \ref{pramplots}. 
For mass accretion coming from the edge,
the ellipsoid is less flattened since the rational support is suppressed by the ram pressure.
There is a lack of solution at high mass because the ram pressure is too large that the system is no longer oblate,
but this is just an unphysical extreme case that our model fails to explain.
As for infall coming along the pole,
the only solution exists for a fully flattened disc.
The radius $R$ of such system is larger compared to that without ram pressure,
but if we evaluate an angle-averaged size, they should be comparable.
With the more realistic estimation, the solution does not change too much with respect to that not considering ram pressure,
except that the range of solution is slightly increased due to extra ram pressure confinement at small mass.
In all cases, the mass-size relation is not very much affected despite of the altered aspect ratio.
Of course the two extreme cases are unrealistic since the focused mass inflow would prevent the gaseous proto-cluster from having an ellipsoidal form.
Nonetheless, they give an idea of the invariability of the mass-size relation with respect to the mass inflow pattern.

\bibliographystyle{aa}
\bibliography{biblio_cluster}

\end{document}